\documentclass[11pt, 5p, nopreprintline]{elsarticle}

\journal{TBA}

\biboptions{numbers,sort&compress}

\usepackage{libertine}
\usepackage{libertinust1math}
\usepackage{amsmath}                 
\usepackage{bbold}                   
\usepackage{graphicx}                
\usepackage{eurosym}                 
\usepackage{mathtools}               
\usepackage{url}                     
\usepackage{booktabs}                
\usepackage{epstopdf}                
\usepackage{xfrac}                   
\usepackage{tabularx}                
\usepackage{bm}                      
\usepackage{subcaption}              
\usepackage{longtable}               
\usepackage{multirow}                
\usepackage{threeparttable}          
\usepackage{pdflscape}               
\usepackage[svgnames]{xcolor}    	 
\usepackage[export]{adjustbox}       
\usepackage[version=4]{mhchem}       
\usepackage{xurl} 				     
\usepackage[colorlinks]{hyperref}    
\usepackage[parfill]{parskip}        
\usepackage[nameinlink,sort&compress,capitalise,noabbrev]{cleveref} 
\usepackage[leftcaption,raggedright]{sidecap} 
\usepackage[prependcaption,textsize=footnotesize]{todonotes} 
\usepackage{siunitx}                 
\usepackage{csquotes} 				 
\usepackage[acronym, automake, nonumberlist]{glossaries-extra}

\usepackage{float}
\usepackage{lipsum}
\usepackage{mdframed}
\usepackage[resetlabels,labeled]{multibib} 
\newcites{S}{Supplementary References}
\bibliographystyleS{elsarticle-num}

\DeclareSIUnit\year{a}
\DeclareSIUnit{\tco}{t_{\ce{CO2}}}
\DeclareSIUnit{\sieuro}{\mbox{\euro}}

\graphicspath{ 						
	{},
	}

\newcommand{\co}{\ce{CO2}}

\newdefinition{rmk}{Remark}
\newtheorem{thm}{Theorem}

\newtheorem{res}[thm]{Result}  

\makeglossaries

\newacronym{eu}{EU}{European Union}
\newacronym{bdew}{BDEW}{Bundesverband der Energie- und Wasserwirtschaft}
\newacronym{dea}{DEA}{Danish Energy Agency}
\newacronym{jrc}{JRC}{Joint Research Center}
\newacronym{tyndp}{TYNDP}{Ten-Year Network Development Plan}
\newacronym{entsoe}{ENTSO-E}{European Network for Transmission System Operators for Electricity}
\newacronym{entsog}{ENTSO-G}{European Network of Transmission System Operators for Gas}

\newacronym{tes}{TES}{Thermal Energy Storage (in form of water tanks)}
\newacronym{chp}{CHP}{Combined Heat and Power plant}
\newacronym{cop}{COP}{Coefficient of Performance}
\newacronym{PV}{PV}{Solar photovoltaics}
\newacronym{ghg}{GHG}{Greenhouse gas}
\newacronym{helmeth}{HELMETH}{Integrated High-Temperature Electrolysis and Methanation for Effective Power to Gas Conversion}
\newacronym{ocgt}{OCGT}{Open Cycle Gas Turbine}
\newacronym{v2g}{V2G}{Vehicle to Grid}
\newacronym{dsm}{DSM}{Demand Side Management}
\newacronym{hvdc}{HVDC}{High Voltage Direct Current}
\newacronym{phs}{PHS}{Pumped Hydro Storage}
\newacronym{dac}{DAC}{Direct Air Capture}
\newacronym{sng}{SNG}{Synthetic Natural Gas}
\newacronym{res}{RES}{Renewable Energy Sources}
\newacronym{cfe}{CFE}{Carbon-Free Electricity}
\newacronym{ldes}{LDES}{Long Duration Energy Storage}
\newacronym{ccs}{CCS}{Carbon Capture and Sequestration}

\newacronym{fom}{FOM}{Fixed Operation and Maintenance costs}
\newacronym{vom}{VOM}{Variable Operation and Maintenance costs}
\newacronym{capex}{CAPEX}{Capital expenditures}
\newacronym{opex}{OPEX}{Operational expenditures}
\newacronym{ppa}{PPA}{Power Purchase Agreement}
\newacronym{ptc}{PTC}{Production Tax Credit}
\newacronym{necp}{NECP}{National Energy and Climate Plan}
\newacronym{ets}{ETS}{Emissions Trading System}
\newacronym{cfd}{CfD}{Contract for Difference}
\newacronym{go}{GO}{Guarantees of Origin}
\newacronym{rec}{REC}{Renewable Energy Certificate}

\newacronym{pypsa}{PyPSA}{Python for Power System Analysis}
\newacronym{nuts}{NUTS}{Nomenclature of Territorial Units for Statistics}
\newacronym{ci}{C\&I}{Corporate and Industry (sectors)}
\newacronym{ngo}{NGO}{Non-Governmental Organization}

\begin{document}

\begin{frontmatter}

	\title{On the means, costs, and system-level impacts of 24/7 carbon-free energy procurement}

	\author[tubaddress]{Iegor Riepin}
	\ead{iegor.riepin@tu-berlin.de}
	\author[tubaddress]{Tom Brown}
	\ead{t.brown@tu-berlin.de}

	\address[tubaddress]{Department of Digital Transformation in Energy Systems, TU Berlin, Germany}

	\begin{abstract}
		A growing number of public and private energy buyers are interested in 24/7 carbon-free energy (CFE) procurement, which means that every kilowatt-hour of electricity consumption is met by carbon-free sources at all times.
		It has the potential to overcome the limitations of established procurement schemes, such as the temporal mismatch between clean electricity supply and buyers' demand that is inherent to \enquote{volumetric} matching.
		Yet it is unclear how 24/7 CFE procurement affects the rest of the power system, and whether this effect is consistent across regional contexts and different levels of system cleanness.
		We use a mathematical model to systematically examine different designs, optimal procurement strategies, costs, and impacts of the 24/7 CFE matching, both for participating buyers and for regions where voluntary procurement occurs.
		We examine mechanisms driving system-level emissions reduction and how they vary across regions and over time.
		Our results indicate that clean energy procurement commitments have consistent beneficial effects on participants and the electricity system.
		Even as grids become cleaner over time, the hourly matching strategy contributes significantly to system-level emissions reduction.
		In addition, voluntary commitments to 24/7 CFE have a further transformative effect on electricity systems through accelerated innovation and early deployment of advanced energy technologies.
	\end{abstract}



\end{frontmatter}

\tableofcontents

\section{Introduction}
\label{sec:intro}


Climate change is driving a global effort to rapidly decarbonise electricity systems across the globe.
The European Union (\gls{eu}) has committed to achieving net-zero emissions by 2050, and the European Commission has adopted the European Climate Law with a target of reducing net greenhouse gas emissions by at least 55\% by 2030, compared to 1990 levels \cite{EuropeanClimateLaw2020}.
National climate initiatives are gaining ground across the globe, including the United States with the Biden administration's goal of reducing emissions by 50-52\% by 2030, compared to 2005 levels \cite{BidenClimatePlan2021}, and China's commitment to have CO$_2$ emissions peak before 2030 and achieve carbon neutrality before 2060 \cite{ChinaNetZero-IEA}.


Many public and private energy buyers have joined this effort by purchasing clean energy to demonstrate their sustainability credentials.
The established certification schemes for voluntary clean energy procurement are the Guarantees of Origin (\gls{go}s) in Europe and the Renewable Energy Certificates (\gls{rec}s) in the United States.
A common feature of these schemes is that renewable energy credits are \enquote{unbundled} from megawatt-hours of energy and can be traded numerous times on specialized markets before being retired.
The owner gets to claim environmental benefits when \enquote{retiring} (i.e. using) the certificate.
The unbundled renewable energy credits, however, have been criticized for not ensuring a physical match between renewable energy generation and consumption \cite{spglobal-recs, bock-icelandGOproblem, re100report-2020}, and for absence of evidence that such schemes drive build-out of additional renewable energy capacity \cite{bjorn-RECSnatcom-2022, gillenwater-2014}.

Some corporate and industrial (\gls{ci}) buyers recongnised the problems associated with the unbundled certificates and turned towards Power Purchase Agreements (\gls{ppa}s).
When signing a \gls{ppa}, energy buyers pledge to buy both the energy and the environmental credits \enquote{bundled} to megawatt-hours.
Market-driven procurement of renewable energy has become one of the major drivers of the energy transition: \gls{ppa}s are expected to account for one-fifth of utility solar PV and wind capacity expansion in 2023--2024, and almost twice as much (36\%) when China is excluded \cite{iea-REppa2023}.
In some countries, such as Spain and Sweden, market-driven procurement of renewable energy capacity surpasses policy-driven capacity expansion \cite{iea-REppa2023}.

\textbf{The key problem --} When \gls{ci} consumers purchase renewable credits, either via unbundled schemes or via bundled \gls{ppa}s, renewable energy supply is typically matched over a long period of time with buyers' energy demand.
As an example, more than 400 members of the RE100 group \cite{re100report-2020} have committed to purchasing renewable energy that matches 100\% of their electricity consumption on an annual basis.
Even the \enquote{best case} of 100\% annual matching commitments, where energy buyers sign \gls{ppa}s with renewable energy generators located within a local grid, has a major problem: the temporal mismatch between renewable energy supply and electricity demand.
In hours when wind and solar generation is low, energy buyers have to depend on carbon-emitting technologies, like coal and gas, available on local markets.
As a result, the 100\% annual renewable matching approach does not entirely eliminate carbon emissions from the energy buyer's portfolio.
Furthermore, the 100\% annual matching approach keeps electricity buyers exposed to price volatility, as the price of electricity on the wholesale markets can vary significantly from hour to hour.


\textbf{One proposed solution: 24/7 CFE --} There is growing interest from the leaders in the clean energy procurement to move beyond the 100\% annual matching approach.
One approach is called 24/7 Carbon-Free Energy (\gls{cfe}) procurement, which aims to match clean energy supply with electricity demand on an \textit{hourly basis}.
Thus, 24/7~CFE procurement has the potential to eliminate all greenhouse gas emissions associated with electricity use of the participating companies.
Furthermore, 24/7~CFE procurement is focused on \textit{carbon-free} rather than renewable energy, which opens up opportunities for a broad range of advanced energy technologies, such as long-duration energy storage (\gls{ldes}) or new clean firm power generators that are needed for deep decarbonisation of energy systems.

Google led the 24/7 movement in 2020 by announcing its commitment to cover the demand at all its data centres and campuses worldwide using the 24/7 CFE approach by 2030 \cite{google-247by2030}.
Shortly after, Google published a policy roadmap on achieving the goal \cite{google-PolicyRoadmap}.
Also, Microsoft has committed to the 100/100/0 by 2030 goal (i.e. \enquote{100\% of electricity consumption, 100\% of the time, matched by zero carbon energy purchases} \cite{Microsoft-vision}.
The list of entities that have committed to 24/7 CFE procurement includes also utilities, such as Peninsula Clean Energy (a community choice aggregator in California) committed to 24/7~CFE by 2025 \cite{peninsula-OurPathto247}, and even cities like Des Moines, Iowa U.S., with a commitment by 2035 \cite{iowaenvcouncil-247}. In 2021, an international group of energy suppliers, electricity buyers, and \gls{ngo}s launched the 24/7 Carbon-free Energy Compact \cite{gocarbonfree247}.
With over 135 signatories, the group aims to \enquote{develop and scale high-impact technologies, energy policies, procurement practices, aand solutions to make 24/7 Carbon-Free Energy achievable for all}.


\textbf{Existing literature on 24/7 CFE --} The growing interest in the means, costs and the system-level impacts of hourly \gls{cfe} procurement has created a need for a quantitative analysis of the approach.
There is relatively little research on the system-level impacts of 24/7~CFE procurement; and it is all fairly recent.
The first to incorporate it into the electricity system planning model were Xu et al. (2022) \cite{xu-247CFE-SSRN}.\footnote{A working paper updates a prior report published in November 2021 investigating the system-level impacts of 24/7~CFE in two regions of the U.S. -- California and PJM \cite{xu-247CFE-report}.}
The authors study the system-level impacts of 24/7~CFE procurement based on an example of California in the year 2030.
In November 2022, IEA also released a model-based study on the 24/7~CFE, focusing on India and Indonesia \cite{ieaAdvancingDecarbonisationClean2022}.
In early 2023, Peninsula Clean Energy published an own white paper introducing their modeling tool---MATCH (Matching Around-the-Clock Hourly energy)---and the results of the modeling indicating feasibility of the company's \enquote{24/7 renewable energy by 2025} goal \cite{peninsula-report247}.

The first studies indicate that voluntary 24/7~CFE commitments reduce the emissions of participating buyers as well as emissions in the electricity grids they operate.
What has been missing is a comprehensive analysis of 24/7~CFE procurement over multiple regions and years, since electricity system characteristics, National Climate and Energy Plans (\gls{necp}s), renewable resource quality, and the cost of clean energy vary significantly across regions and over time.
Furthermore, it is unclear whether decarbonisation effect of 24/7 CFE procurement remains relevant when electricity grids are progressively decarbonised, as is the case in many countries and jurisdictions, or if the 24/7~CFE procurement is only impactful in the early stages of the energy transition.


\textbf{Contribution --} The novelty of this study is to go beyond case studies and show in a systematic way, across many regions and levels of energy transition, how voluntary commitments to 24/7 CFE procurement impact participating buyers and regions where voluntary procurement occurs.
We explore different designs, optimal procurement strategies, costs, benefits and system-level impacts of the 24/7~CFE matching approach.
Our analysis focuses on four regions within the European energy system with unique characteristics, and we vary the analysis years to track the evolution of the energy system.
Additionally, we examine the underlying mechanisms that drive the system-level reduction of CO$_2$ emissions, and examine how they perform over time and across regions.
This work is a follow-up to the earlier public report by the authors \cite{riepin-zenodo-systemlevel247} with an updated model, updated system-wide assumptions, and additional analysis of the results.


The remainder of the paper is structured as follows: \cref{sec:methods} introduces a mathematical model of clean energy procurement strategies and gives a brief summary of the methodology, sources of model input data, and the experimental setup.
The results are presented and discussed in \cref{sec:results} and put in perspective in \cref{sec:discussion}.
The work is concluded in \cref{sec:conclusion}.
\nameref{sec:si} provides further details about model assumptions and parametrisation of the background energy system.

\section{Methods and experimental design}
\label{sec:methods}

\textbf{Optimisation model --} The simulations were carried out with the modified version of the open European power system model PyPSA-Eur, built using the open framework PyPSA \cite{brownPyPSAPythonPower2018}.
The objective of the model is to minimise the annualised system-wide costs of meeting energy demand within the modelled system, while adhering to all relevant engineering, reliability, and policy constraints.
We use a brownfield investment approach, which means that we keep the existing power generation and storage assets in the model and only optimise the capacity expansion of new assets.
The optimal solution of the model includes the location and capacity of new power generation and storage assets, as well as the hourly dispatch of all existing and new assets, including the linear optimal power flow on the transmission network for the planning year.
This is standard for the PyPSA-Eur model \cite{horschPyPSAEurOpenOptimisation2018} and the long-term energy planning models of this class \cite{jenkinsGenX2022, howellsOSeMOSYSOpenSource2011}.

\textbf{Clean energy procurement model --} We introduce new model components (parameters, variables, constraints) into the PyPSA-Eur model to represent voluntary clean energy procurement.
By incorporating these into the mathematical problem, we model a situation where a fraction of electricity demand in a particular bidding zone is voluntarily committed to procuring clean energy with a desired strategy.
The fraction is set at 10\% of the total \gls{ci} electricity demand in a particular bidding zone, hereinafter referred to as \enquote{participating consumers}.
The model co-optimises procurement and operational decisions of the participating consumers to meet their electricity demand in accordance with the procurement strategy they have chosen.

We model and compare two procurement strategies:

\begin{itemize}[-]
    \item \textit{100\% annual matching}: participating buyers meet their energy demand on an annual volumetric basis with additional renewable energy procurement;
    \item \textit{24/7 hourly matching}: participating buyers optimise their investment and operational decisions mixing procurement of additional generation and storage resources and electricity imports so that \gls{cfe} supply meets electricity demand 24/7 throughout the year with the desired quality score.
    \item For benchmarking, we also model \textit{a reference case}: participating buyers have no voluntary environmental commitments and purchase all their electricity from the local grid.
\end{itemize}

\textit{100\% annual matching --} The annual volumetric renewable matching strategy is modelled with a constraint (\ref{eqn:RES100}).
More formally, the sum of all dispatch $g_{r,t}$ of contracted renewable generators $r\in RES$ over the year $t\in T$ is equal to the annual electricity demand of participating consumers:
\begin{equation}
\sum_{r\in RES, t\in T} g_{r,t} = \sum_{t\in T} d_t
\label{eqn:RES100}
\end{equation}

The contracted renewable generators must be new, i.e., they must be additional to the system.
Procured generators can be sited only in the local bidding zone; thus, we model the \enquote{best case} of the annual matching strategy.

\textit{24/7 CFE hourly matching --} The hourly matching strategy is modelled with a constraint (\ref{eqn:CFE}). The 24/7~CFE procurement framework is based on the methodologies and metrics paper by Google~(2021) \cite{google-methodologies}. The framework considers various \gls{cfe} supply options for the participating consumers, including procurement of their own generation resources and storage assets, and electricity imports from the local grid.

More formally, the hourly generation from the procured CFE generators $r\in CFE$, discharge and charge from the procured storage technologies $s\in STO$, plus imports of electricity from the regional grid $im_t$ multiplied by the grid's hourly CFE factor $CFE_t$ minus the excess of the CFE supply must be higher or equal than a certain CFE score $x$ multiplied by the total load of participating consumers:

\begin{equation}
    \begin{split}
        \sum_{r\in CFE, t\in T} g_{r,t} &+ \sum_{s\in STO, t\in T} \left(\bar{g}_{s,t} - \underline{g}_{s,t}\right) \\
        &- \sum_{t\in T} ex_t + \sum_{t\in T} CFE_t \cdot im_t \geq x \cdot \sum_{t\in T} d_t
    \end{split}
\label{eqn:CFE}
\end{equation}

\noindent where \textit{CFE score} $x$[\%] measures the degree to which hourly electricity consumption is matched with carbon-free electricity generation.
Equation (\ref{eqn:CFE}) thus allows for controlling \textit{the quality score} of the 24/7~CFE procurement by adjusting the parameter $x$. The best quality score---100\% CFE---means that every kilowatt-hour of electricity consumption is met by carbon-free sources at all times.

Similarly to the hourly matching strategy, the contracted generators must be additional to the system and can be sited only in the local bidding zone.

Note that if total electricity generation of assets procured by participating consumers exceeds demand in a given hour, the \enquote{excess} carbon-free electricity is not counted toward a CFE score.
Here we assume that the excess can either be curtailed or sold to the regional electricity market at wholesale market prices.
We set a constraint on the total amount of excess generation sold to the regional grid, setting the limit to 20\% of the total annual demand of participating consumers.\footnote{In \nameref{sec:discussion}, we explain the rationale behind this limit and discuss the implications of this assumption.}
The wholesale market prices are derived from dual variables of a nodal energy balance constraint.

The CFE factor of the regional grid ($CFE_t$) can be seen as the percentage of clean electricity in each MWh of imported electricity to supply demand of participating consumers in a given hour.
To compute $CFE_t$, we consider both the hourly electricity mix in the local bidding zone and the emissionality of imported electricity from the neighbouring zones.
The methodology to calculate the grid CFE factor is described in the prior work of the authors \cite{riepin-zenodo-systemlevel247}.

Note that in equation (\ref{eqn:CFE}), $CFE_t$ is affected by additional CFE resources procured by the participating consumers.
This introduces a nonconvex term to the optimisation problem.
The nonconvexity can be avoided by treating the grid CFE factor as a parameter that is iteratively updated (starting with $CFE_t =0 \,~\forall t$).
We find that one forward pass (i.e. 2 iterations) yields very good convergence.

\textbf{Model scope --} Geographically, the model encompasses the entire \gls{entsoe} area, which covers the entire European electricity system.\footnote{Islands that are not connected to the main European system, such as Malta, Crete and Cyprus, are excluded from the model.}
In this study, electricity demand, supply and power transmission infrastructure (alternating current lines at and above 220~kV voltage level and all high voltage direct current lines) are clustered to 37 individual bidding zones \cite{PyPSAEur-docs-spatialresolution}.
Thus, each zone represents a country; some countries straddling different synchronous areas are split into individual bidding zones, such as DK1 (West) and DK2 (East).

Participating consumers can be located in either of the four selected zones: Ireland, Denmark (zone DK1), Germany, or Poland.
The comparison of results across these zones allows us to generalize the implications of voluntary clean energy procurement, taking into account factors such as electricity demand, weather, availability of renewable resources, existing electricity generation capacity mix, and national energy policies, among other factors.

We model two individual years: 2025 and 2030.
From a modelling perspective, a five-year step changes many system parameters.
In particular, technology costs decline as a result of economies of scale and incremental innovation, \gls{necp}s become tighter, \co~prices increase, and some legacy power plants go out of business.
The problem is solved with a temporal resolution of 2920 snapshots, representing a 3-hourly average of the hourly time-series data.

\textbf{Carbon-free energy technologies in scope --} Depending on the scenario, participating consumers can choose from a variety of energy technologies available on the European market. By grouping technologies according to their maturity levels and availability on the market, we formulate three scenarios:

\begin{itemize}[-]
    \item \textit{Palette 1} -- technologies that are commercially available today: onshore wind, utility scale solar photovoltaics (\gls{PV}), and battery storage;
    \item \textit{Palette 2} -- all above plus a long-duration energy storage (\gls{ldes}) that is expected to be commercially available in the near future;
    \item \textit{Palette 3} -- all above plus promising advanced technologies that are in prototype stage today, but are expected to be commercially available in the future: (i) Allam cycle power plant with \gls{ccs}\footnote{Allam cycle is a natural gas power plant with up to 100\% of carbon capture and sequestration} to represent a technology option with higher operational expences (\gls{opex}), and (ii) \enquote{an advanced dispatchable generator}, a stand-in for clean firm technologies with higher capital expences (\gls{capex}), such as advanced geothermal or nuclear systems.
\end{itemize}

\textbf{Technology assumptions --} Cost and other assumptions for energy technologies available for participating consumers were collected primarily from the Danish Energy Agency \cite{DEA-technologydata} for the respective years and are provided in [\cref{tab:tech_costs}].
For the \gls{ldes}, we assume an underground hydrogen storage in salt caverns with an electolyzer and a fuel cell for hydrogen conversion.
Data for advanced clean firm technologies is less reliable due to technological uncertainty and lack of commercial experience; therefore, use our own indicative assumptions.

\textbf{Other assumptions --} Other model inputs and key background system assumptions are provided in [\labelcref{sec:si_2}]. For a full list of technology assumptions, see the GitHub repository \cite{github-247CFEpaper}.

\section{Results}
\label{sec:results}

Throughout the analysis, we examine how voluntary clean electricity procurement strategies affect both buyers and the electricity system as a whole.
From the \gls{ci} buyers' perspective, we examine optimal procurement strategies (i.e., a mix of locally procured energy generation and storage assets and electricity imports), the average cost of procured electricity and the carbon emissions associated with electricity usage, also known as Scope 2 emissions \cite{GHGProtocolScope2}.
From a system perspective, we analyze how voluntary clean energy procurement impacts CO$_2$ emissions in the region and the \enquote{demand pull} for advanced energy technologies.

Our experimental setup incorporates a wide range of scenarios, allowing us to generalise our results beyond individual assumptions.
For convenience, we start with a \textit{base scenario} presented in \cref{subsec:base}.
The base scenario reflects a case when 10\% (on volumetric basis) of \gls{ci} consumers in \textit{Ireland} in \textit{2025} are committed to procuring additional clean energy with a desired strategy.
The participating consumers can attain the 24/7~CFE procurement targets by co-optimizing local procurement using mature technologies, designated as \textit{palette~1}, and by importing electricity from the local grid.
Afterward, we delve deeper into the scenario space, altering selected parameters to explore various alternatives and address the following questions:

\begin{enumerate}[-]
\item \cref{subsec:palette}: What if a wider palette of carbon-free technologies were available to participating buyers?
\item \cref{subsec:location}: What if participating buyers are located in areas of the European electricity grid that differ significantly from the base scenario?
\item \cref{subsec:time}: What are the implications if we move the timeframe ahead five years to 2030?
\item Finally, in \cref{subsec:mechanisms}, we get a closer look at how 24/7 CFE commitments can help decarbonize local grids and reveal the individual mechanisms involved.
\end{enumerate}

Considering the large scenario space, this work presents results for a selection of scenario branches that generalise the system impacts of the 24/7 carbon-free energy procurement.
We publish all code to reproduce the experiments under open license alongside the paper.
See \nameref{sec:code}.

\subsection{Base scenario}
\label{subsec:base}

\begin{res}
    Through 24/7 CFE procurement, participating consumers can eliminate carbon emissions associated with their electricity consumption and help to decarbonise the entire electricity system.
\end{res}

\begin{res}
    90\%--95\% 24/7 CFE targets have only a small cost premium over 100\% annual renewable matching. 100\%~CFE target can be achieved with solar PV, wind and battery storage, but at a high cost premium for participating consumers.
\end{res}

\begin{figure}
    \centering
    \includegraphics[width=0.95\columnwidth]{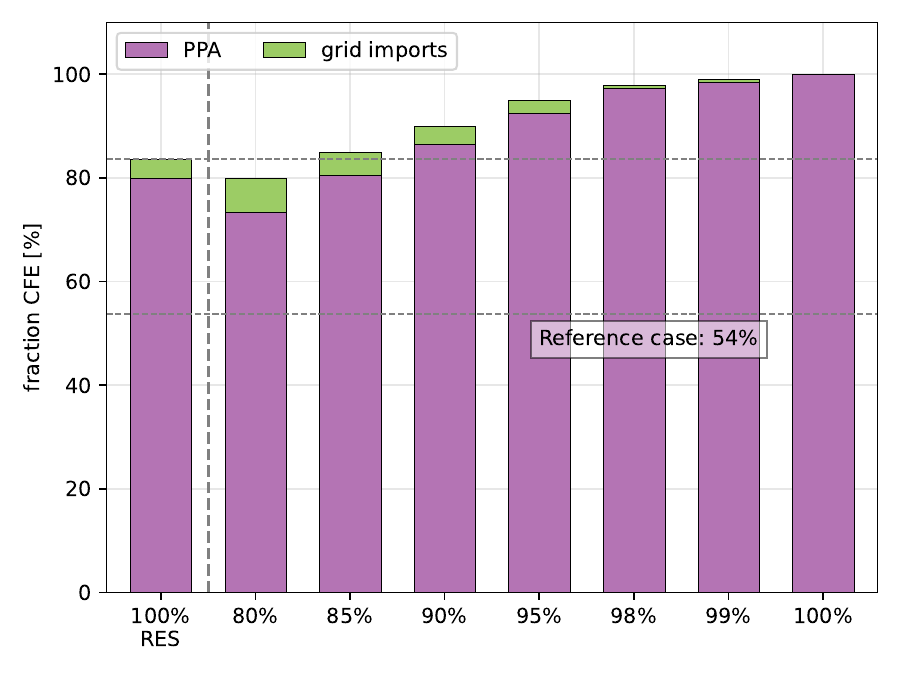}
    \caption{Fraction of hourly demand met with carbon-free electricity.}
    \label{fig:10-2025-IE-p1-used}
\end{figure}

\cref{fig:10-2025-IE-p1-used} shows the fraction of hourly demand met with \gls{cfe} depending on the procurement policy that participating consumers follow.
We contrast two main cases, one where participants commit to 100\% annual renewable matching policy (hereinafter -- \enquote{100\% RES}) and another case where the consumers commit to 24/7~CFE hourly matching policy (hereinafter -- \enquote{24/7~CFE}) with various CFE targets in a range from 80\% to 100\%.
In the \textit{reference case}, participating consumers cover their load purely with grid purchases without any policy regarding the origin of electricity.

In the reference case, only 54\% of demand is met with \gls{cfe}.
100\% RES---the best case for the annual renewable matching policy---results in a 83\% fraction.
Consequently, CFE targets beyond 85\% yield higher share of hourly demand met with \gls{cfe} than 100\%~RES.
Another notable trend in \cref{fig:10-2025-IE-p1-used} is that participating consumers rely more on procured resources and less on grid imports with higher CFE targets.
When targets approach 100\%, participating consumers cannot rely on electricity imports from the grid, since the grid supply mix contains some carbon content in almost all hours.

\begin{figure}[t]
    \centering
    \begin{subfigure}[t]{0.95\columnwidth}
        \centering
        \caption{Average emissions rate of participating consumers.}
        \includegraphics[width=0.95\columnwidth]{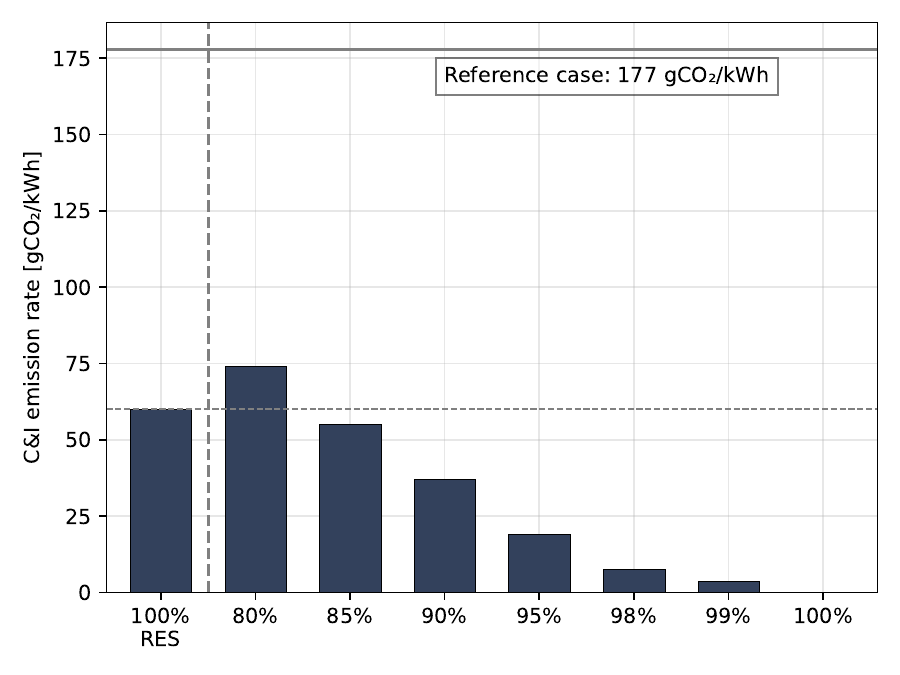}
        \label{fig:10-2025-IE-p1-ci_emisrate}
    \end{subfigure}
    \begin{subfigure}[t]{0.95\columnwidth}
        \centering
        \vspace{-0.2cm}
        \caption{CO$_2$ emissions in the local region (Ireland).}
        \includegraphics[width=0.95\columnwidth]{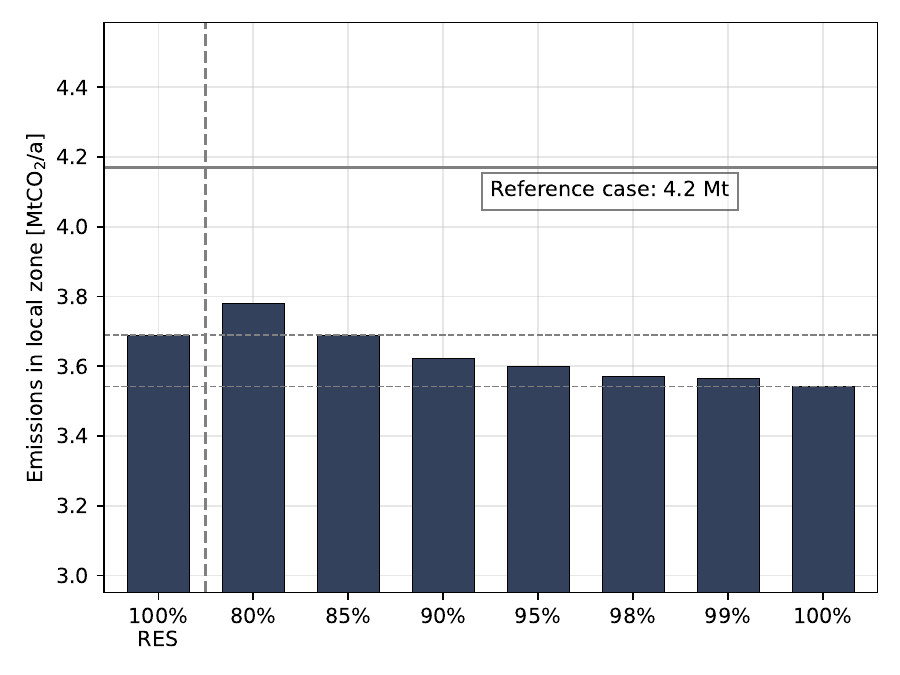}
        \label{fig:10-2025-IE-p1-zone_emissions}
    \end{subfigure}
    \caption{Impact of clean electricity procurement commitments on emissions of participating consumers (top panel) and the system-level emissions in a local region (bottom panel).}
    \label{fig:10-2025-IE-p1-emissions}
\end{figure}

\cref{fig:10-2025-IE-p1-ci_emisrate} shows how choice of a procurement policy affects the average emissions rate of participating consumers.
The average emissions rate is derived by dividing the total CO$_2$ emissions linked to the electricity consumption of participating consumers by the total kilowatt-hours (kWh) consumed.
Already in 2025, Ireland has a moderately clean electricity system: the emission rate is at 177~g\co/kWh in the reference case.
If participating consumers follow 100\%~RES policy, the emission rate is reduced to 60~g\co/kWh.
This yields a considerable emission rate reduction compared to the reference case.
The 24/7~CFE procurement with targets above 85\% helps the participating consumers to reduce emissions further beyond the treshhold set by the 100\%~RES.
At the 100\% CFE target, participating consumers attain a zero emission rate associated with their electricity consumption, as they perfectly align demand and \gls{cfe} supply on an hourly basis.

Voluntary procurement of clean electricity also helps decarbonise the electricity system as a whole.
We explore this effect by plotting CO$_2$ emissions in the local region of the participating consumers, i.e. Ireland (\cref{fig:10-2025-IE-p1-zone_emissions}).
Without voluntary procurement, the model estimates Irish power sector carbon emissions to be at the level of 4.2 Mt\co/a.\footnote{For comparison, \href{https://www.seai.ie/data-and-insights/seai-statistics/key-publications/co2-emissions-report/}{seai.ie} reports this value to be at 10.3~MtCO$_2$ in 2021. A strong decreasing trend is expected, since the Irish government has set ambitious targets to achieve a low-carbon energy system \cite{seaiCOEmissionsReport2020}.}
The modelling results indicate that a commitment to 100\% 24/7~CFE in 2025 from 10\% of the \gls{ci} electricity demand could yield a reduction in Irish emissions by 0.6~Mt\co/a, in contrast to a scenario absent voluntary procurement.
This reduction equates to 14\% of the total emissions from the Irish power sector.
100\%~RES can also facilitate system-level CO$_2$ emissions reductions (as depicted by the left bar in \cref{fig:10-2025-IE-p1-zone_emissions}); however, 24/7~CFE attains superior emissions reductions when the CFE target surpasses 85\%.

\begin{figure}[t]
    \centering
    \begin{subfigure}[t]{0.95\columnwidth}
        \centering
        \caption{Portfolio capacity procured by participating consumers.}
        \includegraphics[width=0.95\columnwidth]{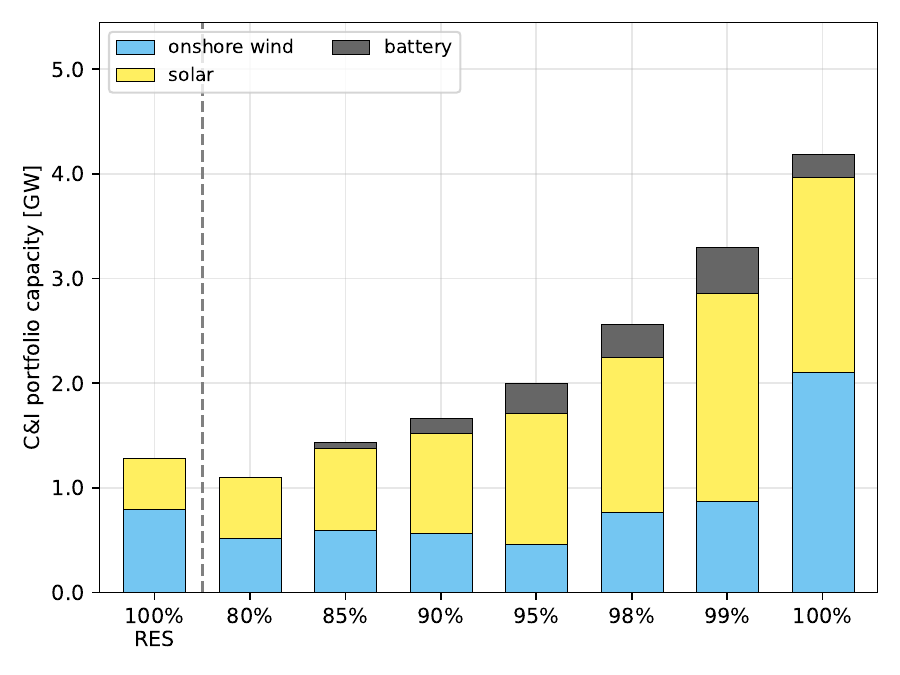}
        \label{fig:10-2025-IE-p1-ci_capacity}
    \end{subfigure}
    \begin{subfigure}[t]{0.95\columnwidth}
        \centering
        \vspace{-0.2cm}
        \caption{Cost breakdown of a procurement policy.}
        \includegraphics[width=0.95\columnwidth]{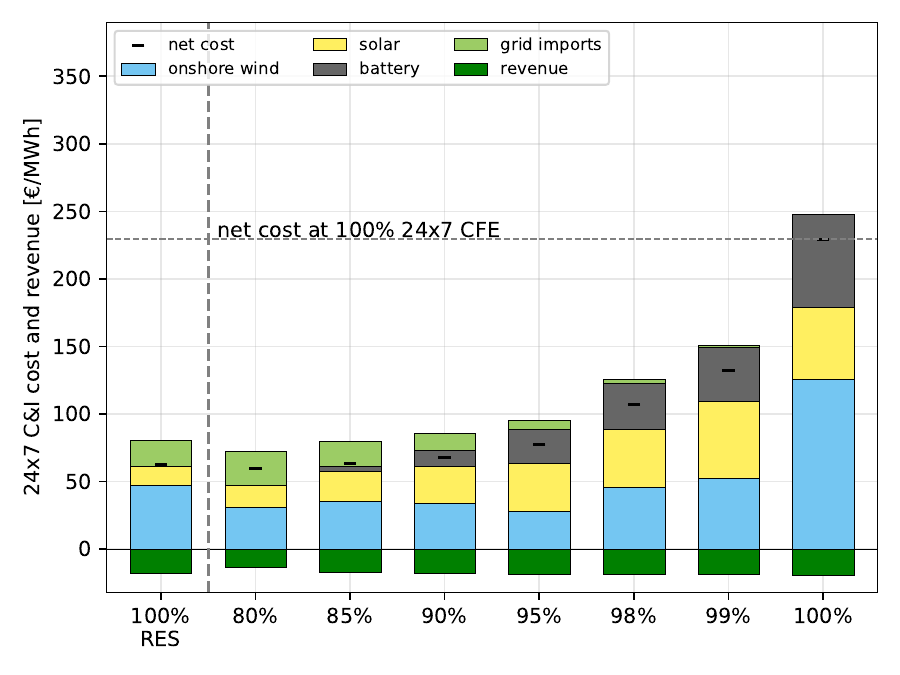}
        \label{fig:10-2025-IE-p1-ci_costandrev}
    \end{subfigure}
    \caption{Optimal capacity portfolios (top panel) and costs breakdown (bottom panel) per procurement strategy.}
    \label{fig:10-2025-IE-p1-ci_procurement}
\end{figure}

The following analysis unveils the modelled cost-optimal procurement strategies for each procurement policy (\cref{fig:10-2025-IE-p1-ci_capacity}).
In Ireland, 10\% of the \gls{ci} sector's demand yields a load of 220~MW.
The findings indicate that, to cover this load adhering to a 100\%~RES policy, participating consumers procure a combined capacity of 1.3~GW from onshore wind and solar PV generators.
Ensuring demand is matched with \gls{cfe} on an hourly basis requires a considerably more substantial portfolio of wind and solar PV, up to 4.0~GW of combined capacity to reach the 100\%~CFE target.
Batteries are integrated into the cost-optimal portfolio mix when CFE targets exceed 85\%.
It is noteworthy that, for an 80\%~CFE target, participating consumers procure less capacity than under a 100\%~RES policy, as they rely more on grid imports.

The breakdown of costs associated with a procurement policy that participating consumers choose is shown in \cref{fig:10-2025-IE-p1-ci_costandrev}.
Note that revenues from selling the excess electricity to the regional grid at market prices can be treated as \enquote{negative costs} and subtracted from the net procurement cost.
The results show that a CFE targets of 80\%--95\% can be achieved at a small cost premium to 100\%~RES policy with solar, wind and batteries.
However, what stands out in the plot is the rapid increase of procurement costs for high 24/7 CFE targets.
For example, 98\% CFE target has cost premium of 54\% over 100\% the annual matching policy; while the last 2\% of hourly matching more than doubles the costs.

\subsection{Technology access}
\label{subsec:palette}

\begin{res}
    24/7 CFE procurement creates an early market for advanced energy technologies the system will need later: long-duration energy storage and clean firm generators.
\end{res}

\begin{res}
    The cost premium of 24/7 procurement with high CFE targets may substantially decrease if advanced energy technologies are available for commercial scale-up.
\end{res}

The cost premium of 24/7~CFE procurement presented in \cref{subsec:base} is driven by the variability of renewable power supply.
Indeed, in the periods when not much wind or solar is available, matching every kWh of electricity consumption with carbon-free electricity on hourly basis is not an easy task.
Battery storage technology is cost-optimal for shifting surplus power supply by a couple of hours.
However, bridging extended periods of low wind and sun with battery storage is expensive.
Also, the results above show that 24/7 participating consumers rely little on grid supply to reach CFE targets of 98\%--100\%, since a local grid's electricity mix almost always includes some fossil-fueled generation.

\begin{figure*}
    \centering
    \begin{subfigure}{0.5\textwidth}
        \centering
        \caption{Cost breakdown of a procurement policy (w/ \gls{ldes}).}
        \includegraphics[width=0.95\columnwidth]{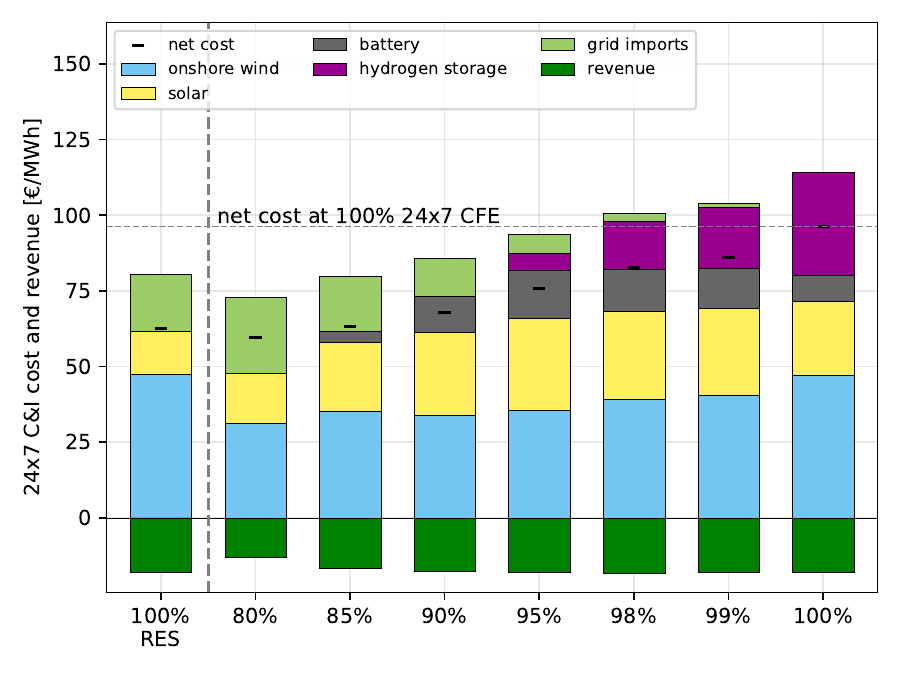}
        \label{fig:10-2025-IE-p2-ci_costandrev}
    \end{subfigure}%
    \begin{subfigure}{0.5\textwidth}
        \centering
        \caption{Costs of a procurement policy \\
        (w/ \gls{ldes} and advanced technologies).}
        \includegraphics[width=0.95\columnwidth]{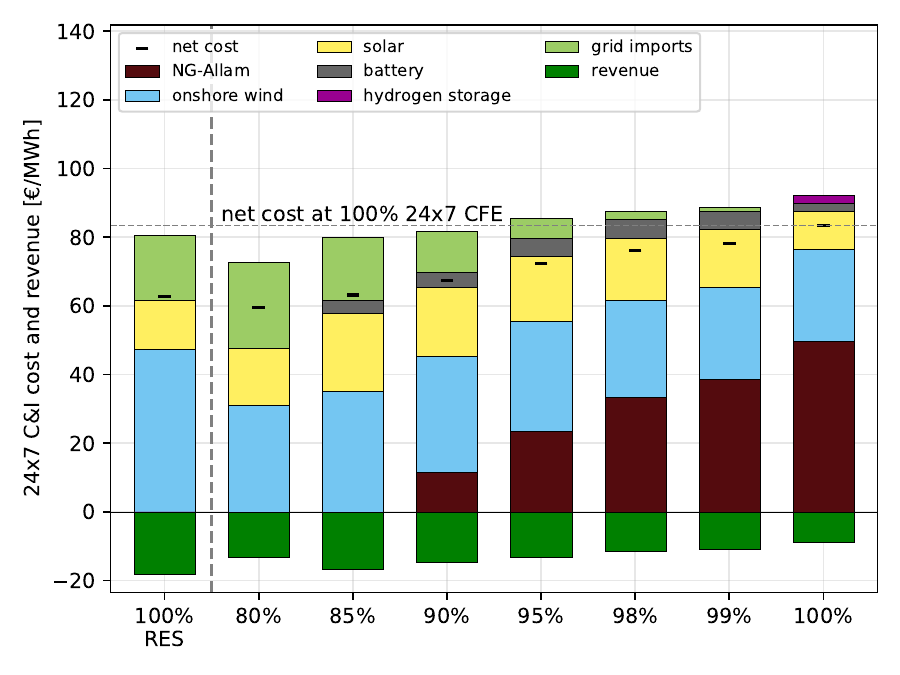}
        \label{fig:10-2025-IE-p3-ci_costandrev}
    \end{subfigure}
    \caption{The breakdown of costs per procurement policy if participating consumers have an access to a wider palette of technologies: w/ \gls{ldes} (left panel); w/ \gls{ldes} and advanced clean firm generators (right panel).
    }
    \label{fig:10-2025-IE-p23-ci_procurement}
\end{figure*}

The results in \cref{fig:10-2025-IE-p23-ci_procurement} reflect a case when 24/7~CFE participants have access to a wider palette of technologies, including \gls{ldes} and advanced clean firm generators, such as advanced geothermal systems, advanced nuclear or Allam cycle generators with \gls{ccs}.

\cref{fig:10-2025-IE-p2-ci_costandrev} (left panel) shows the breakdown of costs for a case when participating consumers have an access to palette~2 technologies, i.e., long-duration storage complements onshore wind, solar PV and battery storage.
Here we assume an underground hydrogen storage in salt caverns with energy storage cost of 2.5~\euro/kWh [\cref{tab:tech_costs}].
An \gls{ldes} system helps to bridge hours with no renewable feed-in.
As a result, the required portfolio of renewable capacity and procurement costs are reduced, especially for high CFE targets.
Thus, 100\% 24/7~CFE policy has a cost premium of 42\% above 100\%~RES, which is significantly lower than without \gls{ldes}.

\cref{fig:10-2025-IE-p3-ci_costandrev} (right panel) depicts the technological palette~3 scenario, wherein participating consumers additionally gain access to clean firm generation technologies.
The Allam cycle turbine with \gls{ccs} now enters into the optimal procurement strategy.
The cost-optimal share of to clean firm technology increases as CFE targets rise.
The clean dispatchable generator helps to fill in the gaps in variable renewable generation to match demand and \gls{cfe} supply on an hourly basis.
The cost premium is further reduced: 100\% 24/7 CFE costs only 15\% above the 100\%~RES.

\subsection{Diverse grid locations}
\label{subsec:location}

\begin{res}
    Despite the inherent variability in regional charac\-teristics---ranging from renewable resources and national policies to degrees of interconnection---the model shows a consistent effects of 24/7 clean energy procurement on both participating buyers and overarching system-level impacts across all analyzed regions.
\end{res}

While the primary focus has been a base scenario exemplifying Ireland, analogous modelling can be applied to diverse regions within the European electricity grid, each exhibiting unique characteristics influenced by local resources, renewable potentials, \gls{necp}s, the degree of interconnections, and other factors.

\begin{figure*}
    \centering
    \begin{subfigure}{0.5\textwidth}
        \centering
        \caption{Average emissions rate of participating consumers (Germany).}
        \includegraphics[width=0.95\columnwidth]{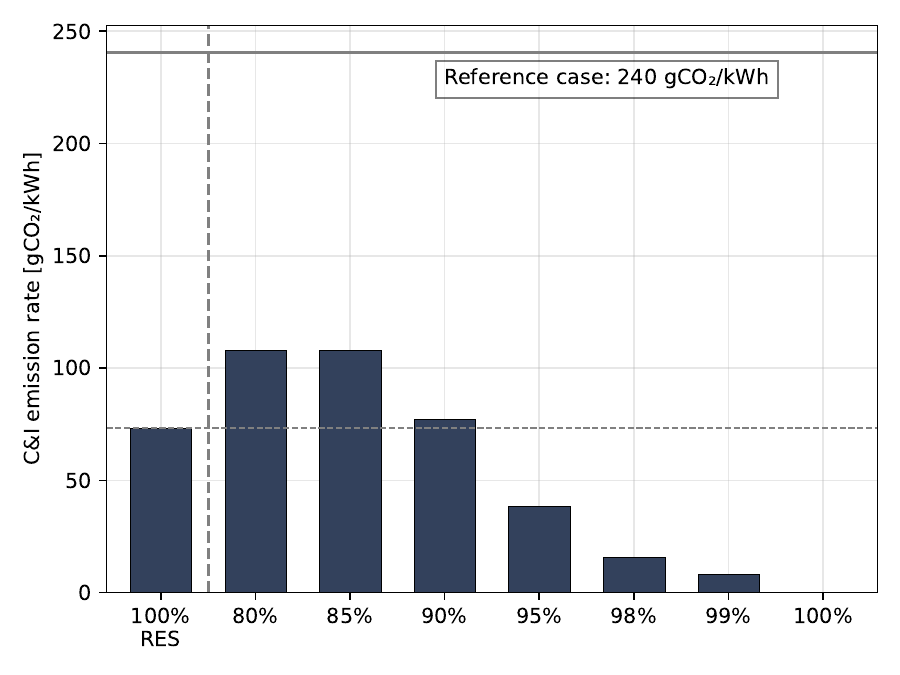}
        \label{fig:10-2025-DE-p3-ci_emisrate}
    \end{subfigure}%
    \begin{subfigure}{0.5\textwidth}
        \centering
        \caption{Average emissions rate of participating consumers (Poland).}
        \includegraphics[width=0.95\columnwidth]{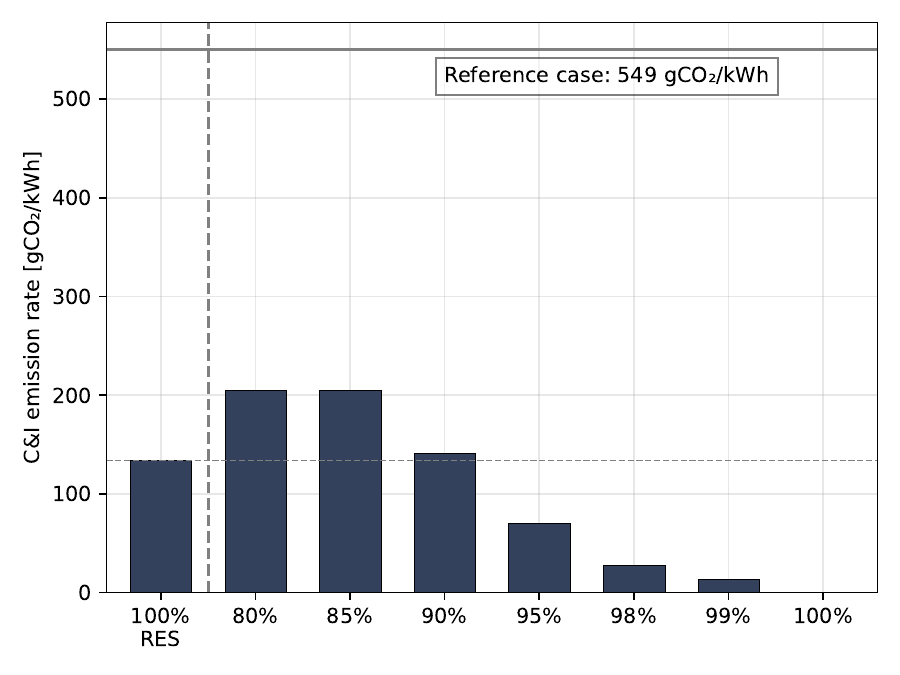}
        \label{fig:10-2025-PL-p3-ci_emisrate}
    \end{subfigure}

    \begin{subfigure}{0.5\textwidth}
        \centering
        \caption{CO$_2$ emissions in the local region (Germany).}
        \includegraphics[width=0.95\columnwidth]{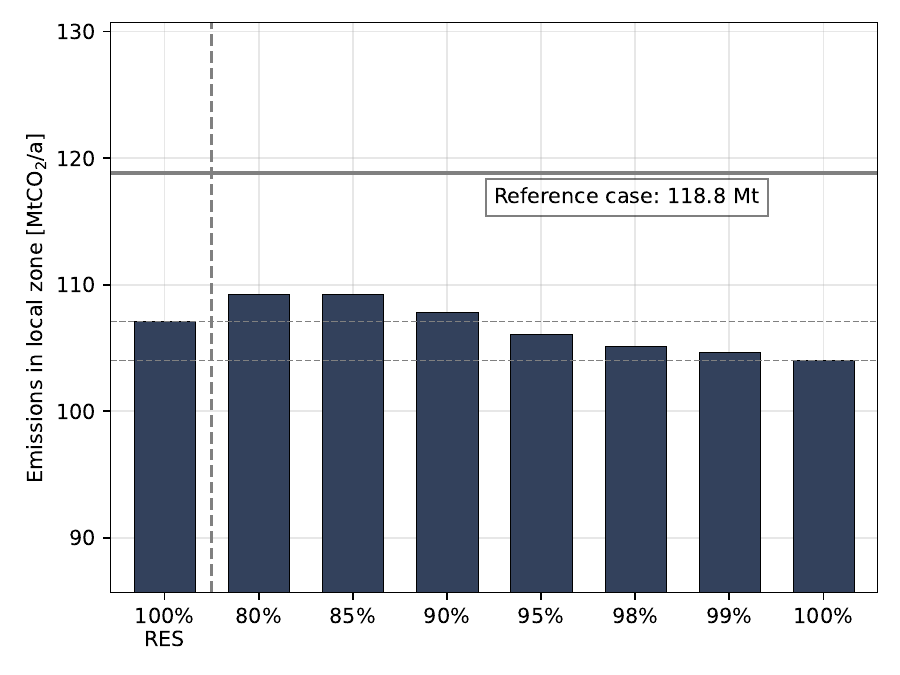}
        \label{fig:10-2025-DE-p3-zone_emissions}
    \end{subfigure}%
    \begin{subfigure}{0.5\textwidth}
        \centering
        \caption{CO$_2$ emissions in the local region (Poland).}
        \includegraphics[width=0.95\columnwidth]{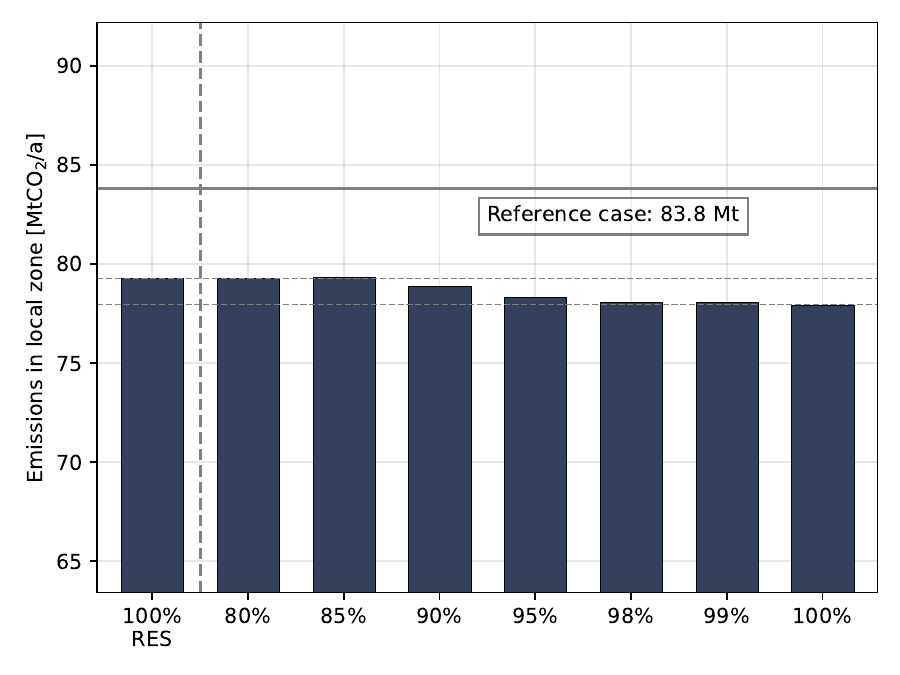}
        \label{fig:10-2025-PL-p3-zone_emissions}
    \end{subfigure}%

    \begin{subfigure}{0.5\textwidth}
        \centering
        \caption{Cost breakdown of a procurement policy (Germany).}
        \includegraphics[width=0.95\columnwidth]{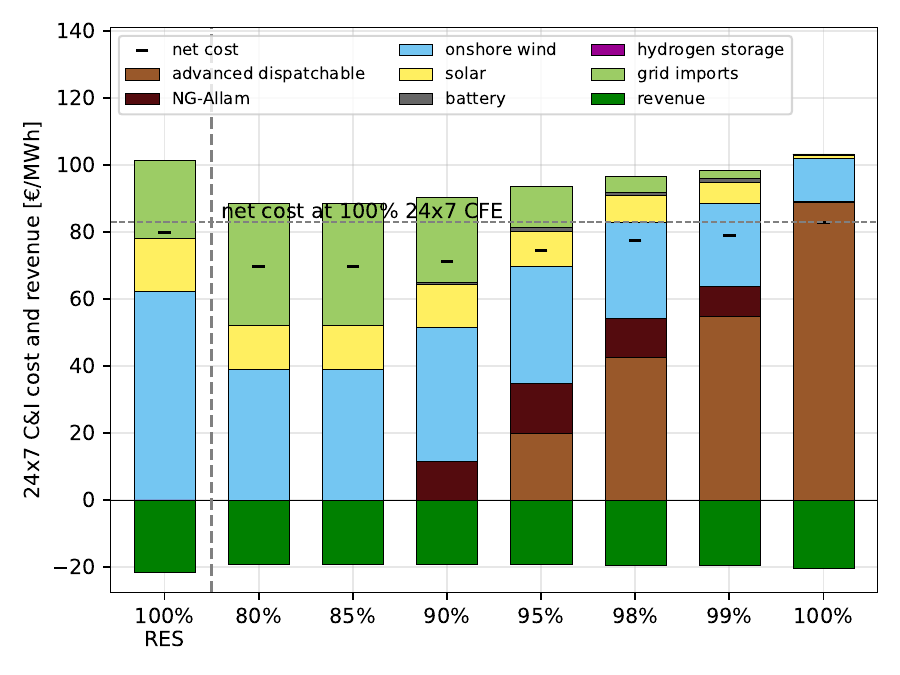}
        \label{fig:10-2025-DE-p3-ci_costandrev}
    \end{subfigure}%
    \begin{subfigure}{0.5\textwidth}
        \centering
        \caption{Cost breakdown of a procurement policy (Poland).}
        \includegraphics[width=0.95\columnwidth]{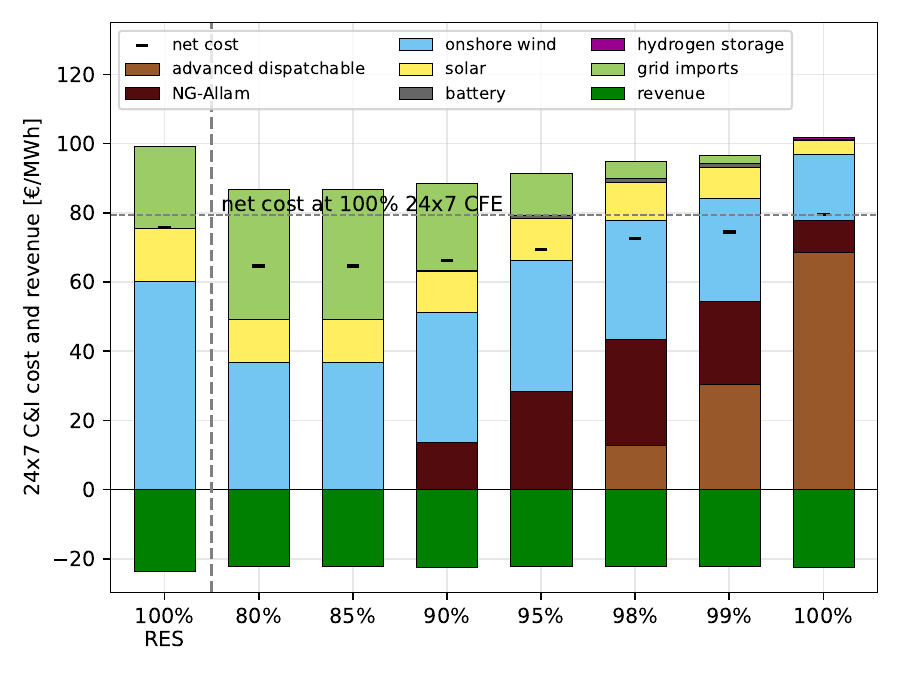}
        \label{fig:10-2025-PL-p3-ci_costandrev}
    \end{subfigure}
    \caption{Selected results for scenarios when participating consumers are located in Germany (left) and Poland (right); all plots are for technological palette~3.}
    \label{fig:10-2025-DEPL-p3-4plots}
\end{figure*}

\cref{fig:10-2025-DEPL-p3-4plots} depicts outcomes for the selected regions of Germany and Poland.
Germany is selected due to its extensive interconnections with neighbouring regions with relatively clean grids, such as France and Denmark, and its status as the largest electricity consumer in Europe.
In contrast, Poland has a smaller electricity demand.
Moreover, the Polish energy system typically has higher carbon intensity, largely due to its historical dependence on coal resources.
The participation rate is maintained at 10\% of the \gls{ci} sector in both regions, aligning with the base scenario.\footnote{A 10\% participation rate within the \gls{ci} sector corresponds to loads of 3.8~GW in Germany and 1.1~GW in Poland.}
Despite regional variances, the dynamics and system impacts of clean energy procurement, as observed in the base scenario, exhibit similarities.

German consumers experience lower average emissions rates, attributed to a cleaner grid, in contrast to those in Poland.
In the reference cases, emission values stand at 240~g\co/kWh and 549~g\co/kWh for Germany and Poland, respectively.\footnote{For context, the emission intensity of electricity generation in 2021 was 402~g\co/kWh for Germany and 750~g\co/kWh for Poland \cite{EEA-GNGEmissionsEU}.}
Echoing the Irish scenario, both \cref{fig:10-2025-DE-p3-ci_emisrate} and \cref{fig:10-2025-PL-p3-ci_emisrate} demonstrate that in both contexts, participating consumers attain reduced emissions rates with 24/7~CFE procurement compared to 100\%~RES, given sufficiently strict CFE targets.

The impact of 24/7 hourly CFE procurement on system-level CO$_2$ emissions is notable in both Germany and Poland, as depicted in \cref{fig:10-2025-DE-p3-zone_emissions} and \cref{fig:10-2025-PL-p3-zone_emissions}.
In the absence of any procurement strategy, carbon emissions from the power sectors of Germany and Poland are projected to be 118.8~MtCO$_2$ and 83.8~MtCO$_2$ in 2025, respectively.%
\footnote{Despite Germany consuming more electricity, the similar scale of power sector carbon emissions between the two countries can be attributed to Poland's significantly higher emission intensity in its electricity mix.}
By engaging in voluntary clean energy procurement, \gls{ci} consumers can drive notable decarbonisation in the local regions, exceeding the benchmarks set by \gls{necp}s.
For instance, in Germany, even a 10\% participation rate under a 100\%~RES policy can diminish national emissions by 11.1~MtCO$_2$ annually; remarkably, the 24/7~CFE approach can amplify this impact to up to 14~Mt\co/a, given a CFE target of 100\%.

The optimal capacities of energy technologies and associated costs are illustrated in \cref{fig:10-2025-DE-p3-ci_costandrev} and \cref{fig:10-2025-PL-p3-ci_costandrev}.
Similar to the Irish scenario, participating consumers complement renewable energy procurement with electricity imports from the grid to meet lower CFE targets.
For the realization of strict CFE targets, both the Allam cycle with \gls{ccs} and advanced clean firm generators are incorporated into the technology mix.
This entry of two advanced technologies into the optimal solution space diverges from the Irish findings, where solely the \gls{opex}-heavy Allam cycle turbine is deemed cost-optimal.
The difference is likely attributable to Germany and Poland being more extensively interconnected with the broader European electricity grid compared to Ireland.
Better interconnection allows exhausting all arbitrage opportunities across temporal and spatial dimensions, and maximising the dispatch value derived from an advanced dispatchable generator with high \gls{capex}.
Mirroring the findings from Ireland, the cost premium of 24/7~CFE matching gets minimal when these advanced technologies are accessible.

\subsection{Advancing five years forward}
\label{subsec:time}

\begin{res}
    Over time, participating consumers reap benefits from the gradual decline in clean technology costs and a progressively cleaner state of the electricity grids, enhancing the affordability of 24/7 CFE procurement.
\end{res}

\begin{res}
    Voluntary commitments to 24/7 CFE procurement retain their significance in enhancing system value, even as grids become cleaner. Analogous to the base scenario, hourly matching commitments facilitate a more profound decarbonisation compared to 100\% annual renewable matching, provided that CFE targets surpass a particular threshold.
\end{res}

\begin{figure*}
    \centering
    \begin{subfigure}{0.5\textwidth}
        \centering
        \caption{Fraction of hourly demand met with carbon-free electricity.}
        \includegraphics[width=0.95\columnwidth]{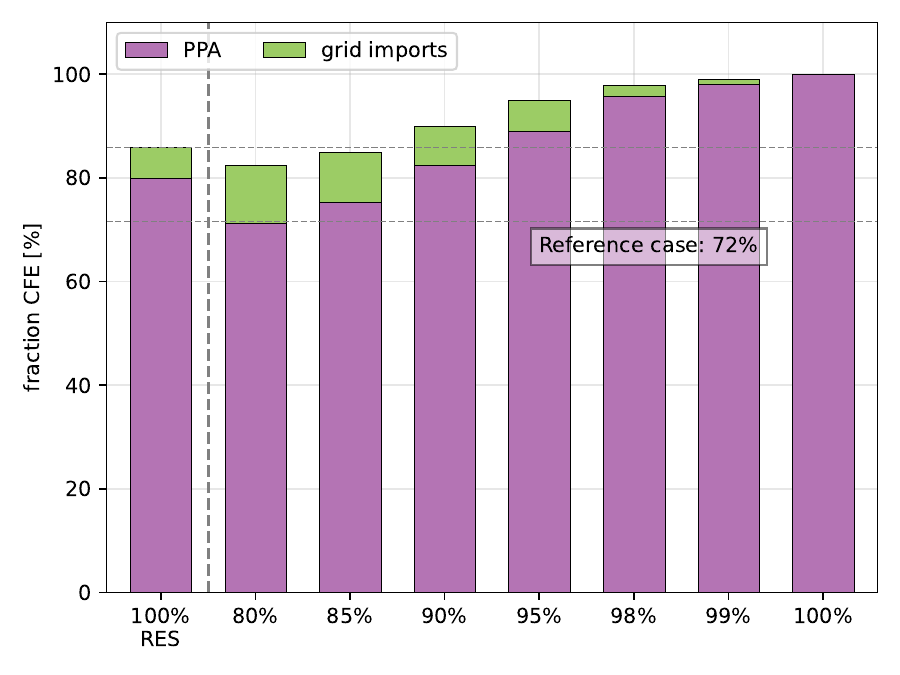}
        \label{fig:10-2030-IE-p1-used}
    \end{subfigure}%
    \begin{subfigure}{0.5\textwidth}
        \centering
        \caption{Average emissions rate of participating consumers.}
        \includegraphics[width=0.95\columnwidth]{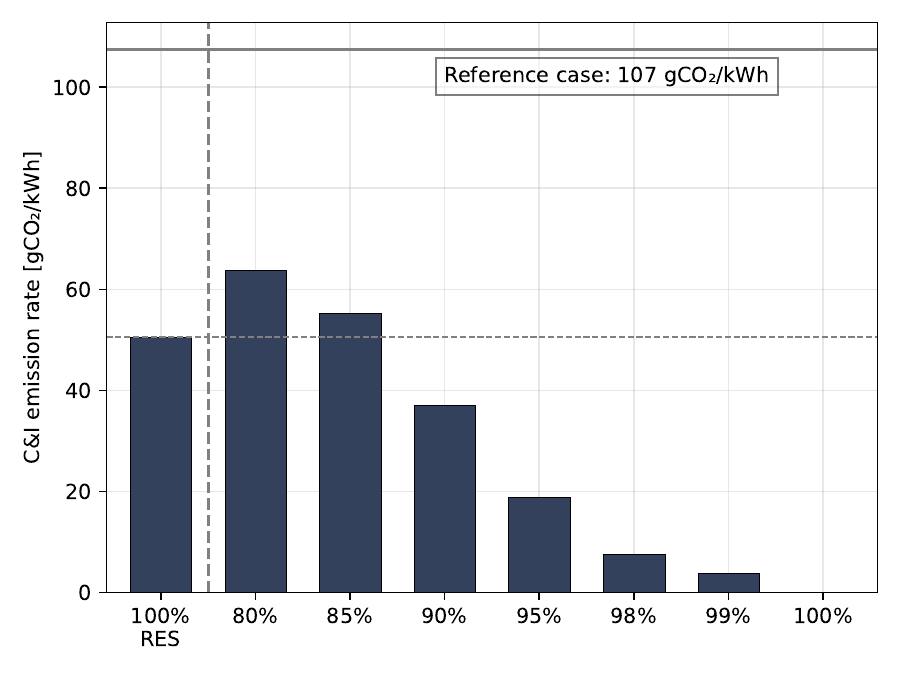}
        \label{fig:10-2030-IE-p1-ci_emisrate}
    \end{subfigure}

    \begin{subfigure}{0.5\textwidth}
        \centering
        \caption{CO$_2$ emissions in the local region (Ireland).}
        \includegraphics[width=0.95\columnwidth]{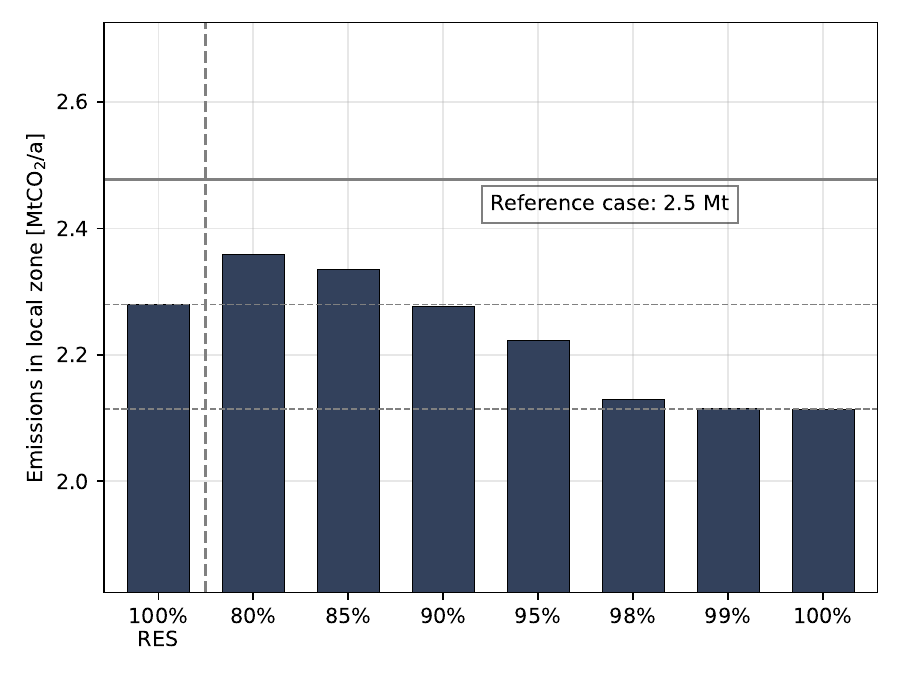}
        \label{fig:10-2030-IE-p1-zone_emissions}
    \end{subfigure}%
    \begin{subfigure}{0.5\textwidth}
        \caption{Cost breakdown of a procurement policy: technological palette~1.}
        \includegraphics[width=0.95\columnwidth]{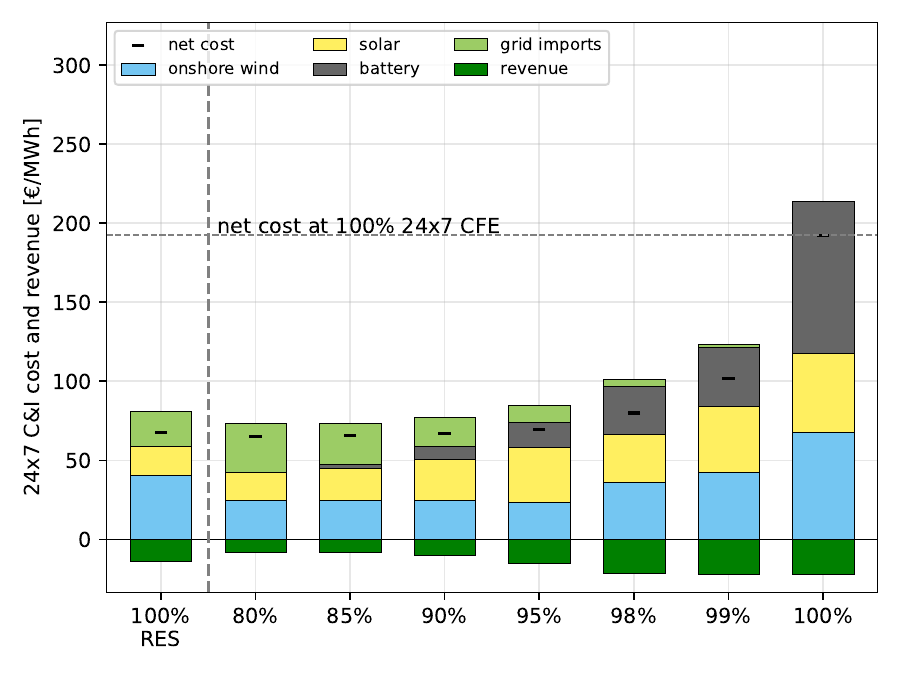}
        \label{fig:10-2030-IE-p1-ci_costandrev}
    \end{subfigure}%

    \begin{subfigure}{0.5\textwidth}
        \centering
        \caption{Cost breakdown of a procurement policy: technological palette~2.}
        \includegraphics[width=0.95\columnwidth]{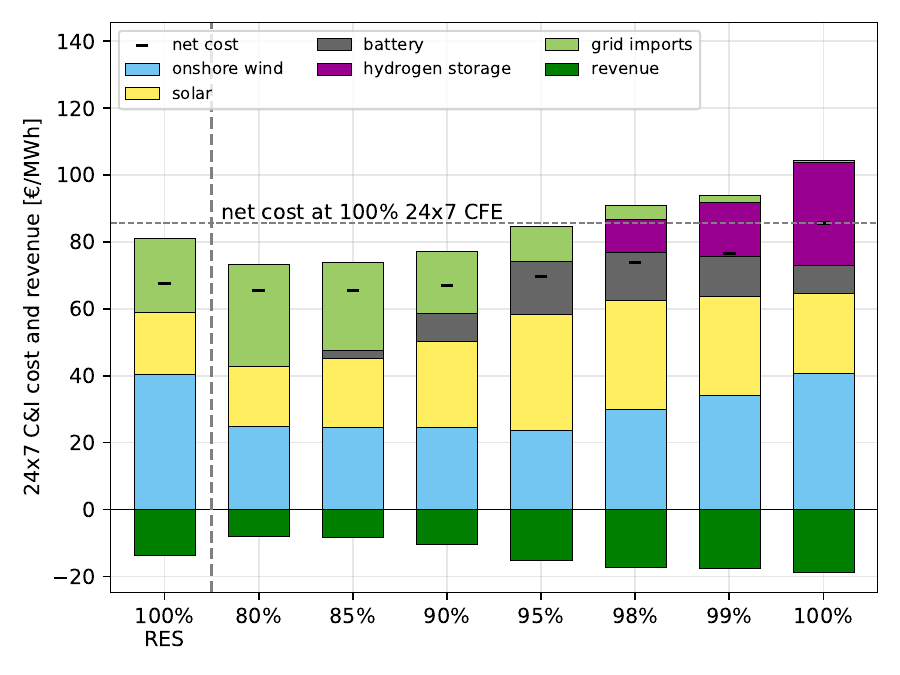}
        \label{fig:10-2030-IE-p2-ci_costandrev}
    \end{subfigure}%
    \begin{subfigure}{0.5\textwidth}
        \centering
        \caption{Cost breakdown of a procurement policy: technological palette~3.}
        \includegraphics[width=0.95\columnwidth]{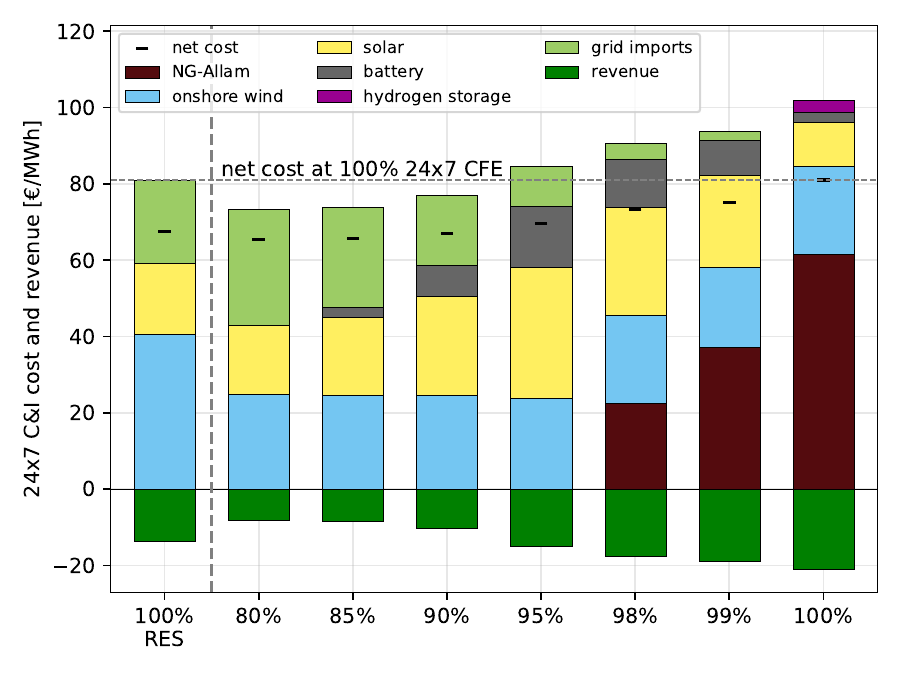}
        \label{fig:10-2030-IE-p3-ci_costandrev}
    \end{subfigure}

    \caption{Results for the scenario of Ireland 2030; 10\% participation rate.
    Figures \ref{fig:10-2030-IE-p1-used}--\ref{fig:10-2030-IE-p1-ci_costandrev} display the technological palette~1 scenario.}
    \label{fig:10-2030-IE-6plots}
\end{figure*}

Having explored the methods, costs, and impacts of clean electricity procurement in the European electricity system for the year 2025 in preceding sections, this section delves into the projected results and implications of these strategies for 2030.
Referencing back to the \nameref{sec:methods} section, the key implications of advancing five years forward are cleaner electricity grids and decreased costs for energy  technologies. \cref{fig:10-2030-IE-6plots} illustrates the outcomes for the \textit{base scenario} transitioned to 2030, maintaining all other parameters constant.

Analyzing \cref{fig:10-2030-IE-p1-used}, which presents the fraction of hourly demand met by clean electricity under each procurement policy in 2030, two key insights emerge when compared to the base scenario.
Firstly, 72\% of demand is met by \gls{cfe} in the reference case, marking an 18\% increase from 2025, indicating an enhancement in the cleanliness of the background electricity grid.
Consequently, lower CFE targets (like 80\%) are met with higher share of electricity imports.
Secondly, participating consumers depend on their own procurement of \gls{cfe} resources and storage, limiting imports from the background grid to achieve strict CFE targets.
This effect stays consistent with the base scenario.

A cleaner background grid results in lower average emissions rates for participating consumers.
\cref{fig:10-2030-IE-p1-ci_emisrate} shows that the emissions rate for consumers purchasing only grid electricity decreases from 240~g\co/kWh in 2025 to 107~g\co/kWh in 2030.
The hourly matching policy enables the attainment of zero emissions linked to \gls{ci} participants' consumption.
It is interesting to note that a CFE target of 90\% yields lower average emissions than a 100\% annual matching policy, even though these policies are equally costly (see \cref{fig:10-2030-IE-p1-ci_costandrev}--\cref{fig:10-2030-IE-p3-ci_costandrev}).

In cleaner systems, both 100\%~RES and 24/7~CFE policies sustain their beneficial impact on system-level emissions.
As observed in \cref{fig:10-2030-IE-p1-zone_emissions}, 24/7~CFE with a sufficiently high CFE target, outperforms 100\%~RES policy in the decarbonisation impact: with 10\% participation rate, hourly matching reduces local emissions by up to 0.4~Mt\(\text{CO}_2\)/a, in contrast to the 0.2~Mt\(\text{CO}_2\)/a reduction achieved by annual matching.
The 0.2~Mt\(\text{CO}_2\)/a differential equates to approximately 8\% of Irish power sector emissions.
The mechanisms driving this impact are detailed in \cref{subsec:mechanisms}.

As illustrated in \cref{fig:10-2030-IE-p1-ci_costandrev}--\cref{fig:10-2030-IE-p3-ci_costandrev}, the cost breakdown of the procurement policy reveals a decline in 24/7~CFE costs across all technological pallets relative to 2025.
Participating consumers benefit from lower prices for clean energy technologies and progressively cleaner electricity grids.
Remarkably, these participants not only achieve zero emissions associated with their consumption and exert a notable impact on system-level \(\text{CO}_2\) emissions but also do so while encountering smaller cost premiums, enhancing the accessibility of the voluntary clean energy procurement goals.

\subsection{Understanding the mechanism of 24/7 CFE procurement in grid decarbonisation}
\label{subsec:mechanisms}

\begin{res}
    24/7 CFE procurement mitigates system-level CO$_2$ \newline
    emissions through two distinct mechanisms: \enquote{profile} and \enquote{volume}. The mechanisms origin from distinct aspects of the interplay between the 24/7 CFE procurement and the background electricity grids.
\end{res}

\begin{res}
    The disparity in decarbonisation outcomes between 24/7 CFE hourly and 100\% annual renewable matching policies becomes increasingly pronounced as local grids transition to cleaner states. This phenomenon underscores the effectiveness of 24/7 CFE procurement in supporting system decarbonisation in the context of evolving national energy and climate policies.
\end{res}

In the previous sections, we demonstrated that 24/7 clean energy procurement results in a substantial reduction of CO$_2$ emissions in local electricity grids (Figures \ref{fig:10-2025-IE-p1-zone_emissions}, \ref{fig:10-2025-DE-p3-zone_emissions}, \ref{fig:10-2025-PL-p3-zone_emissions} and \ref{fig:10-2030-IE-p1-zone_emissions}).
In this section, we examine the underlying mechanisms that drive these observed reductions, so as to better understand how clean energy procurement and emissions reduction interrelate.

As originally identified by Xu et al. (2021) \cite{xu-247CFE-report}, two mechanisms are responsible for reducing system-wide emissions:

First, a \textit{profile} mechanism: participants doing 24/7 hourly matching procure clean energy resources that better match their demand patterns.
As certain consumers align their demand with \gls{cfe} supply on hourly basis, the need for dispatchable generation that can firm intermittent renewable supply is lower in the rest of the system.
This mechanism can reduce the utilisation of fossil-based generators, such as \gls{ocgt} power plants, which typically ramp up when wind and solar resources are scarce.

Second, a \textit{volume} mechanism: even the cost-optimal procurement strategy for 24/7 hourly matching might include some amount of excess clean energy.
If total \gls{cfe} generation of assets procured by 24/7 participants exceeds demand in a given hour, the \enquote{excess CFE} is not counted toward a CFE target; however, it is clean electricity that can potentially be stored (using batteries or LDES), or sold to the regional grid.
As for the latter, excess CFE might displace emitting grid generators or reduce the need to import electricity from neighbouring areas.

Several factors influence the degree to which profile and volume mechanisms contribute to decarbonising the local grids, such as the composition of the 24/7 portfolio, the volume of excess generation sold to the grid, and the electricity generation (and import) mix in the local zone.

We conduct an experiment to isolate the decarbonisation effects of the volume and profile mechanisms.
In the experiment, we reran the optimization while fixing an excess electricity sold to the grid to zero (NB excess is still possible, but must be either stored or curtailed).
Such a model setup includes only a profile decarbonisation mechanism.
We attribute the difference between emissions reduction in the experiment model setup and the base scenario to the volume mechanism.
As a result, the reduction of emissions in a local zone can be attributed either to excess electricity sold to the grid (volume effect) or to better alignment of CFE generation with demand (profile effect).

\begin{figure}[H]
    \centering

    \begin{subfigure}[t]{\columnwidth}
        \centering
        \caption{Emissions reduction in a local zone with profile and volume mechanisms isolated. Data is normalised per MWh of demand participating in 24/7 procurement.}
        \includegraphics[width=0.95\columnwidth]{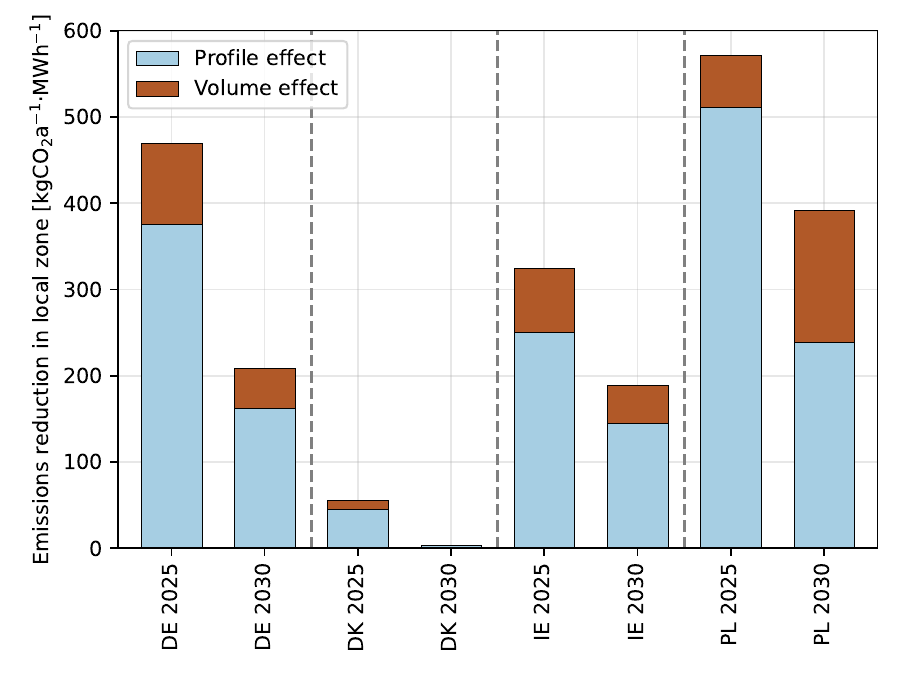}
        \label{fig:10-profile-volume.pdf}
    \end{subfigure}

    \begin{subfigure}[t]{\columnwidth}
        \centering
        \caption{Percentage of emissions reductions in a local zone as a result of a procurement policy compared with the no procurement policy case.}
        \includegraphics[width=0.95\columnwidth]{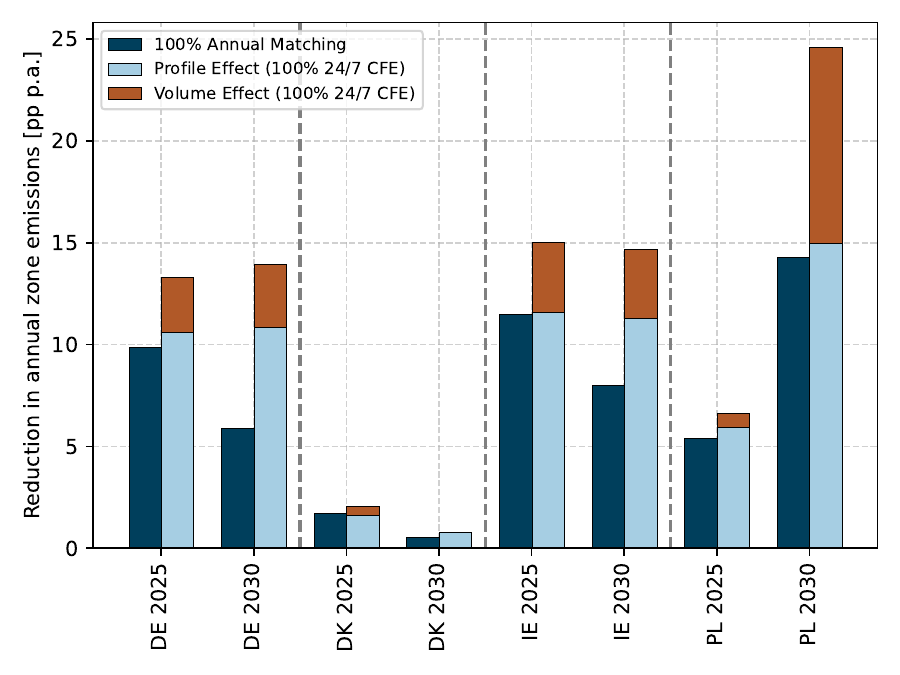}
        \label{fig:10-hourly-annual.pdf}
    \end{subfigure}

    \caption{
        System-level emission reduction: comparison of procurement policies and isolation of decarbonisation mechanisms.
        \cref{fig:10-profile-volume.pdf} shows breakdown of emissions reduction in a local zone if 10\% of \gls{ci} load follows 24/7 hourly procurement with CFE 100\% target. The profile are volume mechanisms are isolated.
        \cref{fig:10-hourly-annual.pdf} compares reductions in annual zone emissions in percent points achieved through 100\% annual renewable matching and 100\% 24/7 CFE matching; no procurement policy is assumed in the counterfactual scenario.}
        \label{fig:decarbonisation_story}

\end{figure}

\cref{fig:10-profile-volume.pdf} illustrates the reduction in local zone emissions attributed to each of the two decarbonisation mechanisms.
In the plot, the x-axis shows the different states of electricity grids in which \gls{ci} consumers participate in 24/7 procurement.
The eight 'background grids' are generated from the combination of modelled wildcards: four zones and two years.

\cref{fig:10-profile-volume.pdf} shows that the impact of 24/7 CFE procurement in absolute terms is proportional to how clean the background grid on average is.
Thus, the largest impact is at 571.6~kgCO$_2$/a·MWh$^{-1}$ takes place in the Poland 2025 zone, and the lowest one is 3.7~kgCO$_2$/a·MWh$^{-1}$ in Denmark 2030 (NB values for 10\% of \gls{ci} sector participation rate with CFE 100\% target).
It is noteworthy that from 2025 to 2030, the decarbonisation effect of voluntary clean electricity procurement decreases over time in all zones since electricity grids become cleaner as a result of national climate policy programs and decomissioning of emitting power plants.
A comparison of the volume and profile mechanisms reveals that both contribute to the decarbonization of the local grids.
In particular, the profile mechanism contributes significantly to overall emissions reductions, with a major share across all zones regardless of their configuration.

Additional insights into decarbonisation mechanisms can be gained by examining the emission reductions achieved through 100\%~RES and 100\% 24/7 CFE matching policies.
\cref{fig:10-hourly-annual.pdf} compares the decarbonisation impact within a local zone of the two procurement policies in relative terms (percent point p.a.), with a counterfactual scenario where no consumers engage in voluntary clean electricity procurement.
The figure demonstrates that, even though the absolute impact of procurement policies diminishes over time, the relative impact of 24/7 CFE on system-level emissions \textit{increases over time} when compared to 100\%~RES.
This phenomenon is attributable to the self-cannibalisation effect of wind and solar that influences the value of additional renewable capacity.
This is relevant in regions where governments adhere to ambitious \gls{necp} targets, like Ireland and Germany (NB Denmark is an outlier with an exceptional RES target exceeding 100\% of national electricity demand for 2030).
The profile mechanism maintains its relevancy even when renewable energy generators undergo self-cannibalization.
The volume mechanism retains, or even amplifies, its relevance, by leveraging a diversified portfolio of procured resources by 24/7 participants, including batteries, long-term energy storage, and advanced clean firm technologies.
Consequently, clean electricity is not only available but can also be sold to the background grid during periods when system-wide wind and solar generation is scarce.

\section{Discussion}
\label{sec:discussion}

\textbf{Our results in perspective --} Our findings agree with the existing model-based studies on clean energy procurement \cite{xu-247CFE-SSRN,ieaAdvancingDecarbonisationClean2022,peninsula-report247}, as well as with the general understanding of the energy research community that \enquote{100\% renewable energy is not enough} \cite{chalendar-2019} in a context of electricity consumers claiming environmental sustainability.
Specifically, we show that through hourly matching of demand with clean electricity, electricity consumers can both completely negate their carbon emissions and contribute to broader system-wide decarbonisation.
100\% CFE matching is viable with commercially available technologies, albeit incurs a cost premium for participating consumers.
The cost premium can be substantially reduced if the advanced technologies, such as long-duration energy storage and clean firm generators are available, or if consumers opt for participating in the 24/7~CFE procurement with quality scores ranging from 90\% to 95\% instead of an absolute 100\%.

Another important finding of this work that confirms prevailing understanding is that 24/7 CFE matching creates an early market for advanced energy technologies. Further research is needed to better understand how 24/7 CFE commitments push advanced energy technologies along their learning curves, and to estimate the potential economic impact on global decarbonisation costs.

According to a comparison of the results over the modeled regions and time periods, voluntary clean energy procurement commitments appear to have consistent effects on both participating buyers and system-level decarbonisation.
Based on these findings, this work can provide a basis for generalising the implications of clean energy procurement strategies, while taking into account regional differences in electricity demand, renewable resource availability, legacy power plant fleets, degrees of interconnection, and energy policies, among other factors.

A deeper look at the role of voluntary clean energy procurement commitments in grid decarbonisation reveals that the hourly matching strategy retains its significance in reducing system-level emissions, even as grids become cleaner over time.
We examined two distinct mechanisms through which 24/7 CFE procurement facilitates system decarbonization: \enquote{profile} and \enquote{volume}.
As companies adapt their procurement strategies, it is important to understand how these mechanisms perform across different grid locations and years.
Interestingly, the disparity between the annual (\enquote{volumetric}) matching with renewable energy and 24/7 CFE hourly matching strategies in decarbonisation outcomes becomes ever more evident as electricity grids become cleaner.
This phenomenon illustrates the effectiveness of 24/7 CFE procurement in supporting system decarbonisation within the context of evolving national energy and climate policies.

\textbf{Features of 24/7~CFE not covered in this work --} Some positive features of 24/7 CFE procurement fall beyond the scope of this manuscript.

From a system-wide perspective, voluntary commitments to 24/7 CFE reduce the need for flexibility in background electricity systems.
There will be less investment in gas-fired peaker generators, as well as less need for battery storage elsewhere in the system.
This effect has been explored by Xu et al. (2022) and Riepin~\&~Brown (2022) \cite{riepin-zenodo-systemlevel247,xu-247CFE-report}

From an electricity buyer perspective, 24/7 CFE procurement can provide a hedge against volatile wholesale market prices.
Since 100\% CFE matching implies covering the entire electricity demand with clean energy from procured generators via power purchase agreements and operating storage, the buyer is not exposed to the wholesale market price risk (see \cref{fig:10-2025-IE-p1-used}).
More information on the price hedge effect is provided by the 24/7 CFE Energy Compact \cite{gocarbonfree247}.

\textbf{Critical appraisal --} This work has simplifications and limitations that should be acknowledged.

First, only the electricity sector is included in the mathematical model of the European energy system, which is justified by the research focus.
However, it is important to acknowledge that the electricity sector is not an isolated system, but is increasingly becoming linked with other energy sectors, such as heat and transport.
Brown et al. (2018) discuss the synergies of sector coupling in the integrated European energy system \cite{brownSynergiesSectorCoupling2018}.

Second, consumers participating in clean energy procurement are modelled as one entity.
The real-world situation may involve a number of individual consumers participating in a voluntary clean energy procurement scheme.
Our assumption is that all consumers that commit to 24/7 matching, form alliances, and enter into contracts with carbon-free energy generators so that their aggregated consumption can be matched with clean generation on an hourly basis to achieve a given CFE target.
In reality, participating consumers can pursue hourly matching strategies independently based on their own load profiles.
See Xu \& Jenkins (2022) investigating this case \cite{princeton-TEACs-2022}.

Third, we assume that the participating consumers have a flat demand profile (i.e. baseload).
As a matter of fact, consumer demand profiles vary based on their business activities.
It has been shown by Riepin~\&~Brown (2022) that the shape of consumption profiles affects the cost-optimal technology mix required for achieving a certain CFE target; however, the shape of the profile has relatively little impact on procurement costs, emissionality of the portfolio, and system-level impacts of 24/7 procurement \cite{riepin-zenodo-systemlevel247}.

Fourth, participating consumers have an inflexible demand, i.e., no demand-side management is considered.
In reality, many \gls{ci} consumers have some degree of demand flexibility, which they can use to adjust their consumption as needed based on grid signals, such as the wholesale market price and the average emission rate of the local electricity mix.
Moreover, some consumers, such as data centers and cloud computing providers, can shift their load also in space, i.e., to shift computing
jobs and associated power loads across data center locations \cite{rosskoningsteinWeNowMore2021}.
In the recent work, Riepin~\&~Brown (2023) explore how space-time load-shifting flexibility can be used to meet high 24/7 CFE targets, as well as what potential benefits it may offer to participating energy buyers and to the rest of the energy system \cite{riepin-zenodo-spacetime247CFE}.

One of the simplifications of mathematical models implementing 24/7 CFE procurement is how the excess carbon-free electricity is handled.
The \enquote{excess CFE} can be defined as the total amount of carbon-free electricity supply exceeding the demand of the participating consumers in a given period of time.
Here we assume that the excess can either be curtailed or sold to the regional electricity market at wholesale market prices, until the excess constraint is reached.
The constraint restricts the volume of excess electricity sold to the regional grid, ensuring it does not surpass 20\% of the annual demand from participating consumers.
The problem with the uncostrained export of excess CFE to the regional market could be a situation where excess electricity from the renewable generators procured by 24/7 consumers displaces electricity from other renewable generators in the market, which in turn may challenge additionality of the 24/7 procurement; and if system-wide renewable expansion is constrained, the cost-optimal solution may be to use the excess clean electricity to \enquote{relax} the constraint, increasing the renewable portfolio of participating consumers.
Similar \enquote{spillover effect} problem has been observed and discussed by Xu et al. (2022) \cite{xu-247CFE-SSRN}.
In their white paper, Peninsula Clean Energy team also discuss the problems and implications of the excess supply. Citing their work: \enquote{We conclude that excess supply is a necessary aspect of a time-coincident renewable portfolio in the range of likely market conditions in California today} \cite{peninsula-report247}.
Both studies perform sensitivity analyses on the excess supply and discuss the implications on their findings \cite{xu-247CFE-SSRN, peninsula-report247}.

\textbf{Is 24/7~CFE the right strategy? --} Last but not least, it is worth mentioning an ongoing discussion whether 24/7 CFE is the right strategy for companies to reduce their carbon emissions.
Some proponents of alternative strategies contend that 24/7 CFE, or \enquote{temporal matching}, may not represent the most effective approach to emissions reduction.
This argument is based on the rationale that optimising procurement and deployment of carbon-free generators and batteries to cover consumption 24/7 with clean electricity is a different strategy to optimising the assets specifically to displace \enquote{as much emissions as possible}.

One of the alternative strategies could be that companies, in their pursuit to reduce system-level carbon emissions, focus on system-wide impact as measured via short-run marginal emissions accounting (\enquote{emissions matching}).
Xu~et~al. (2023) bring this discussion from the realm of arguments into quantitative analysis, concluding that \enquote{emissions matching strategies have zero or near-zero long-run impact on system-level CO$_2$ emissions} \cite{princeton-247CFEvsEmissionality}. The reason for this, citing the authors, is that voluntary carbon-free energy procurement under this strategy almost exclusively displaces other carbon-free energy rather than fossil fuels.

Another alternative strategy could be that companies do annual matching with \enquote{overprocurement}, i.e., covering their own consumption on an annual volumentric basis plus procuring some more clean electricity on top. Such strategy could be implemented at low cost, e.g., with solar PV.
However, as shown in \cref{subsec:mechanisms}, this strategy would not be as effective in reducing system-level emissions as 24/7 CFE matching, especially in the context of evolving national energy and climate policies.
Further, such a strategy would not promote innovation and pull forward in time research and investment decisions in advanced energy technologies, such as long-duration energy storage and clean firm generators.

More work---encompassing the collective contributions of the research community, \gls{ngo}s, policy-makers, and corporate energy buyers---is needed to deepen our understanding of the implications, challenges, and opportunities inherent in various clean energy accounting approaches and procurement strategies, guiding us all towards net-zero sustainable energy systems.

\section{Conclusion and Policy Implications}
\label{sec:conclusion}

In this work, we systematically examined the means, costs, and the system-level impacts of the 24/7 carbon-free energy matching.
We find that 24/7 matching is a viable and effective strategy for electricity buyers aiming to eliminate their own carbon footprint and contribute to wider system decarbonisation.
Voluntary commitments to the hourly matching strategy have a further transformative effect on electricity systems through accelerated innovation and early deployment of advanced energy technologies.
The evidence holds across geographical locations and is robust to sensitivity analyses.
Even as grids become cleaner over time, the 24/7 matching strategy retains its significance in reducing system-level emissions, which is essential in light of evolving national energy and climate policies.

\section*{Acknowledgements}

We thank the following people for fruitful discussions on the 24/7 carbon-free energy procurement and the art of energy system modelling: Fabian Neumann, Elisabeth Zeyen, Devon Swezey, Brian Denvir, Marc Oman, Jesse Jenkins and research staff of the Zero carbon Energy systems Research and Optimization Laboratory (ZERO Lab) at Princeton University.

Iegor Riepin was supported by the research grant from Google LLC.

This work is a follow-up to the prior public report by the authors \cite{riepin-zenodo-systemlevel247} with an updated model, updated system-wide assumptions, and additional analysis of the results.

\section*{Author Contributions}


\textbf{Iegor Riepin}:
Conceptualization --
Data curation --
Formal Analysis --
Investigation --
Methodology --
Software --
Validation --
Visualization --
Writing - original draft

\textbf{Tom Brown}:
Conceptualization --
Methodology --
Software --
Supervision --
Writing - review \& editing

\section*{Data and Code Availability}
\label{sec:code}

The simulations were carried out with PyPSA --- an open-source software framework for simulating and optimising modern energy systems \cite{brownPyPSAPythonPower2018}.

All code, input data and results are published under an open license. The code to reproduce the experiments is available at GitHub \cite{github-247CFEpaper}.

\printglossary[type=\acronymtype]

\addcontentsline{toc}{section}{References}
\renewcommand{\ttdefault}{\sfdefault}
\bibliography{manuscript}


\newpage

\makeatletter
\renewcommand \thesection{S\@arabic\c@section}
\renewcommand\thetable{S\@arabic\c@table}
\renewcommand \thefigure{S\@arabic\c@figure}
\makeatother
\renewcommand{\citenumfont}[1]{S#1}
\setcounter{equation}{0}
\setcounter{figure}{0}
\setcounter{table}{0}
\setcounter{section}{0}

\section*{Supplementary Information}
\label{sec:si}

\section{Technology assumptions}
\label{sec:si_1}

Cost and other assumptions for energy technologies available for participating consumers were collected primarily from the Danish Energy Agency \cite{DEA-technologydata}. These assumptions are listed in \cref{tab:tech_costs}.
A full list of technology assumptions, including the data for energy technologies in the background energy system, is available via the reproducible scientific workflow in the GitHub repository \cite{github-247CFEpaper}.

\begin{table*}[]
    \centering
    \resizebox{0.95\textwidth}{!}{%
        \begin{tabular}{lccccccc}
            \hline\hline
            \textbf{Technology} &
            \textbf{Year} &
            \textbf{\begin{tabular}[c]{@{}c@{}}CAPEX\\ (overnight cost)\end{tabular}} &
            \textbf{\begin{tabular}[c]{@{}c@{}}FOM\\ {[}\%/year{]}\end{tabular}} &
            \textbf{\begin{tabular}[c]{@{}c@{}}VOM\\ {[}Eur/MWh{]}\end{tabular}} &
            \textbf{\begin{tabular}[c]{@{}c@{}}Efficiency\\ {[}per unit{]}\end{tabular}} &
            \textbf{\begin{tabular}[c]{@{}c@{}}Lifetime\\ {[}years{]}\end{tabular}} &
            \textbf{Source} \\ \hline\hline
            Utility solar PV & 2025 & 612 \officialeuro/kW & 1.7 & 0.01 & - & 37.5 & \cite{DEA-technologydata} \\
            & 2030 & 492 \officialeuro/kW & 2.0 & 0.01 & - & 40.0 & \cite{DEA-technologydata}  \\ \hline
            Onshore wind & 2025 & 1077 \officialeuro/kW & 1.2 & 0.015 & - & 28.5 & \cite{DEA-technologydata} \\
            & 2030 & 1035 \officialeuro/kW & 1.2 & 0.015 & - & 30 & \cite{DEA-technologydata} \\ \hline
            Battery storage & 2025 & 187 \officialeuro/kWh & - & - & - & 22.5 & \cite{DEA-technologydata}  \\
            & 2030 & 142 \officialeuro/kWh & - & - & - & 25 & \cite{DEA-technologydata} \\ \hline
            Battery inverter & 2025 & 215 \officialeuro/kW & 0.3 & - & 0.96 & 10 & \cite{DEA-technologydata} \\
            & 2030 & 160 \officialeuro/kW & 0.3 & - & 0.96 & 10 & \cite{DEA-technologydata} \\ \hline
            Hydrogen storage$^1$ & 2025 & 2.5 \officialeuro/kWh$_{\text{H}_2}$ & - & - & - & 100 & \cite{DEA-technologydata}  \\
            & 2030 & 2.0 \officialeuro/kWh$_{\text{H}_2}$ & - & - & - & 100 & \cite{DEA-technologydata} \\ \hline
            Hydrogen electrolysis & 2025 & 550 \officialeuro/kW$_{\text{el}}$ & 2.0 & - & 0.67 & 27.5 & \cite{DEA-technologydata} \\
            & 2030 & 450 \officialeuro/kW$_{\text{el}}$ & 2.0 & - & 0.68 & 30 & \cite{DEA-technologydata} \\ \hline
            Hydrogen fuel cell & 2025 & 1200 \officialeuro/kW  & 5.0 & - & 0.50 & 10.0 & \cite{DEA-technologydata} \\
            & 2030 & 1100 \officialeuro/kW & 5.0 & - & 0.50 & 10.0 & \cite{DEA-technologydata} \\ \hline
            Allam cycle generator$^2$ & 2025 & 2760 \officialeuro/kW  & 14.8 & 3.2 & 0.54 & 30 & \cite{navigant-report, NetZeroAmerica-report} \\
            & 2030 & 2600 \officialeuro/kW & 14.8 & 3.2 & 0.54 & 30 & \cite{navigant-report, NetZeroAmerica-report} \\ \hline
            Clean firm generator$^3$ & 2025 & 10000 \officialeuro/kW & 0 & 0 & 1.00 & 30.0 & [-]\\
            & 2030 & 10000 \officialeuro/kW & 0 & 0 & 1.00 & 30.0 & [-]\\ \hline \hline
        \end{tabular}%
    }
\begin{tablenotes}
    {\footnotesize
    \item[] Notes: All costs are in 2020 euros; $^1$assumed to be underground hydrogen storage in salt caverns; $^2$costs also include estimate of 40 \officialeuro/t\co for carbon transport and sequestration; $^3$a stand-in for clean dispatchable technologies, such as advanced geothermal or advanced nuclear systems; own indicative assumptions. \item[] CAPEX = capital expenditure; FOM = fixed operations and maintenance costs; VOM = variable operations and maintenance costs.
    }
\end{tablenotes}
    \vspace{0.2cm}
    \caption{Technology assumptions.}
    \label{tab:tech_costs}
\end{table*}

\section{Background system assumptions}
\label{sec:si_2}


The model performs \textit{a perfect-foresight optimisation} of investment and power dispatch decisions to meet electricity demand of the 24/7 consumers, as well as the demand of other consumers in the European electricity system for 2025 or 2030.

In all the modelled countries, renewable generation \textit{must meet the political targets} as defined in the \gls{necp}s or by more recent national policy targets (such as the Easter package in Germany). In countries that not have a target for 2025, a linear increase from targets between 2020 and 2030 is assumed. The assumed renewable targets are available in the config file of the scientific workflow \cite{github-247CFEpaper}.

We assume 2013 to be a representative climate year for renewable generation. The renewable feed-in profiles are computed with the atlite package \cite{atlite-github}.

The country demand profiles for 2025 and 2030 are assumed to be the same as in 2013. Time-series data for electricity demand is based on the Open Power System Data project \cite{OPSD}.

The existing power plant fleet data is based on the powerplantmatching package \cite{Powerplantmatching-github}. We consider national policies and decommissioning plans for coal and nuclear power plants based on data from the \enquote{beyondfossilfuels.org} and the \enquote{world-nuclear.org} projects \cite{FossilFuels, WorldNuclearAssociation}.

We assume price for EU ETS allowances to be 80~\euro/t\co and 130~\euro/t\co  for 2025 and 2030, accordingly.
The price for natural gas is assumed to be 35~\euro/MWh.

\end{document}